\documentclass[journal,comsoc]{IEEEtran}
\usepackage[T1]{fontenc}% optional T1 font encoding

\ifCLASSINFOpdf
\else
\fi
\usepackage{algorithm,amsbsy,amsmath,amssymb,epsfig,bbm,mathrsfs,multirow,amsthm,xcolor}
\usepackage{graphicx,caption,subcaption}%for three figures side-by-side
\usepackage{setspace}
\usepackage{algorithmic}
\usepackage{subcaption,float}
\usepackage[hyphens]{url}
\usepackage{cite}
\newtheorem{definition}{Definition}

\usepackage[cmintegrals]{newtxmath}

\usepackage{setspace}

\usepackage{tabularx}
\makeatletter
\newcommand{\multiline}[1]{%
	\begin{tabularx}{\dimexpr\linewidth-\ALG@thistlm}[t]{@{}X@{}}
		#1
	\end{tabularx}
}
\makeatother
\begin{document}
%\raggedbottom
\renewcommand{\baselinestretch}{1}
\title{Joint Speed Control and Energy Replenishment Optimization for UAV-assisted IoT Data Collection with Deep Reinforcement Transfer Learning}

\author{Nam~H.~Chu, Dinh~Thai~Hoang, Diep~N.~Nguyen, Nguyen~Van~Huynh, and Eryk~Dutkiewicz
	
	\IEEEcompsocitemizethanks{\IEEEcompsocthanksitem The authors are with the
		School of Electrical and Data Engineering, University of Technology Sydney, Australia (e-mails: namhoai.chu@student.uts.edu.au, hoang.dinh@uts.edu.au, diep.nguyen@uts.edu.au, huynh.nguyenvan@student.uts.edu.au, and eryk.dutkiewicz@uts.edu.au).
	\IEEEcompsocthanksitem Preliminary results in this paper were presented at the IEEE WCNC Conference, 2021~\cite{chu_fast_2021}}
	}% <-this % stops a space
%\markboth{IEEE Transactions on Communications}%
%{Submitted paper}

\maketitle

\renewcommand{\baselinestretch}{1}	
\begin{abstract}
Unmanned aerial vehicle (UAV)-assisted data collection has been emerging as a prominent application due to its flexibility, mobility, and low operational cost. 
	However, under the dynamic and uncertainty of IoT data collection and energy replenishment processes, optimizing the performance for UAV collectors is a very challenging task. 
	Thus, this paper introduces a novel framework that jointly optimizes the flying speed and energy replenishment for each UAV to significantly improve the data collection performance. 
	Specifically, we first develop a Markov decision process to help the UAV automatically and dynamically make optimal decisions under the dynamics and uncertainties of the environment. 
	We then propose a highly-effective reinforcement learning algorithm leveraging deep Q-learning, double deep Q-learning, and a deep dueling neural network architecture to quickly obtain the UAV's optimal policy. 
	The core ideas of this algorithm are to estimate the state values and action advantages separately and simultaneously and to employ double estimators for estimating the action values. 
	Thus, these proposed techniques can stabilize the learning process and effectively address the overestimation problem of conventional Q-learning algorithms.
	To further reduce the learning time as well as significantly improve learning quality, we develop advanced transfer learning techniques to allow UAVs to ``share'' and ``transfer'' learning knowledge. Extensive simulations demonstrate that our proposed solution can improve the average data collection performance of the system up to 200\% compared with those of current methods.
\end{abstract}


\IEEEpeerreviewmaketitle

%--------------------------------------------------------------------------------------------------------------------------	%--------------------------------------------------------------------------------------------------------------------------
\renewcommand{\baselinestretch}{1.533}	
\section{Introduction}
\label{sec:Introduction}
Over the last ten years, the Internet of Things (IoT) has been rapidly evolving to meet ever-increasing demands in industries and many aspects of our daily lives.
	According to~\cite{Cisco}, the IoT's share in wireless connections will increase from $33\%$ in 2018 to $50\%$ by 2023, which is equivalent to 14.7 billion connections.
	Consequently, the IoT system scale has been growing exponentially,	posing various challenges to service providers and network operators.
	First, the uneven distribution of IoT devices over large areas makes traditional wireless Access Points (APs) (e.g., WiFi APs) inefficient in collecting IoT data. 
	Second, traditional APs are even unable to collect IoT data in various situations.
	For example, IoT devices are distributed at hard-to-reach locations such as on top of trees, outside high-rise buildings, or even under bridges.  
	In this case, cellular base stations can act as data collectors in IoT data collection networks.
	However, the limited energy storage capacity and communication capability hinder IoT devices from transmitting their data over long distances.
	Thus, IoT data collection networks have been demanding effective and flexible solutions for data acquisition.

Recently, UAVs have been emerging as a promising approach to tackle the above challenges.
	In particular, when UAVs act as on-demand flying APs, thanks to their aerial superiority, they can establish good line-of-sight (LoS) links for the IoT nodes.
	This leads to better wireless communication channels, and thereby improving the quality of service (QoS) in comparison with traditional approaches, especially for IoT applications that are sensitive to delay and/or require stable communications~\cite{Wang2019Joint}. 
	In addition, in remote areas without access to terrestrial infrastructures, UAVs can provide a much more economic solution to collect IoT data than traditional approaches, e.g., long-range ground broadcasting stations or high-cost satellite communications.
	Another important advantage of UAVs is that they can be promptly established in emergency circumstances, in which the existing infrastructure is disrupted and incapable of receiving data from IoT devices~\cite{Liu2020Distributed}. 
	Due to the flexibility, mobility, and low operational cost, UAVs have been being deployed as flying APs for some real-world projects, e.g., Google's Loon and Facebook's Aquila~\cite{google_loon, facebook_connecting_2014}.

However, there are still some challenges that hinder the applications of UAVs in IoT data collection networks. 
	In particular, unlike traditional solutions for collecting IoT data (e.g., deploying fixed APs), UAVs have limited energy resources supplied by batteries.
	When the UAVs' batteries are depleted, they must replenish their energy by flying back to the charging stations to charge or replace their batteries. 
	It is worth noting that given a fixed working duration, the more time the energy replenishment process takes, the less time the UAVs can spend for collecting IoT data.
	Alternatively, the energy replenishment process is highly dynamic since it depends on the distance between the UAV and the charging stations. % but also the speed that the UAVs choose to fly back to the charging stations.		
	Therefore, optimizing energy usage and the energy replenishment process is critical to achieving high system performance, but very challenging in practice. 
	Moreover, the UAVs often fly around to collect IoT data, while IoT nodes are statically allocated over different zones, and their sensing data are random depending on surrounding environments. 
		To that end, optimizing operations of UAVs in different zones to maximize data collection efficiency is another major challenge that needs to be considered.
	
	To address the aforementioned problems, jointly optimizing the UAV's speed and energy replenishment process is an effective approach to maximize data collection efficiency. 
		In particular, it is stemmed from the fact that IoT device densities and amounts of sensing data may be different in 
		different zones. 
%		may have different IoT device densities and different amounts of sensing data. 
		Thus, the UAV needs to adjust its speeds over these zones to maximize the data collection efficiency. 
		For example, the UAV may want to fly slowly in a zone with a high IoT device density to increase opportunities of successfully collecting data. 
		Conversely, at other zones that contain less or even no IoT nodes, the UAV can fly faster to avoid missing opportunities from other active zones.			 
		Note that given a fixed flight time, a high speed may cost the UAV more energy than a low speed~\cite{shan_looking_2020}. 
		Therefore, controlling the UAV's speed can also affect the UAV's energy consumption.
		On the other hand, the energy replenishment process influences the system performance from a different angle. 
		In particular, if the UAV can appropriately determine when and where to return to charge/replace its battery, the replenishing energy duration may decrease significantly, leading to an improvement of the system performance.
		Thus, jointly optimizing the UAV's speed and energy replenishment process can maximize the data collection efficiency and conserve more energy. %resulting in the improvement of system performance for UAV-assisted IoT data collection networks.
	
\subsection{Related Works}
\subsubsection{UAV's Energy Replenishment Process} 
In~\cite{zeng_age_2020}, the authors consider a UAV-assisted IoT system where a charging station is deployed to prolong UAV serving time. 
	To minimize the age of information (AoI) under the constraint of the UAV's charging rate and battery capacity, they propose a least-charging-timed Metropolis-Hasting trajectory and a least-visit-time-based trajectory. 
	The authors point out that the UAV's charging rate has much more influence on the low bound of AoI than the UAV's battery capacity. 		
	A similar model is proposed in~\cite{bouhamed_uav_2020}. However, in this work, the authors aim to minimize the data collection time by optimizing the UAV trajectory and the order of IoT devices that the UAV is going to visit. Alternatively, they also employ a deep deterministic policy gradient algorithm to help the UAV learn an optimal trajectory. Furthermore, to speed up the training process, a transfer learning model is introduced.
	However, the simulation results do not show the benefit of applying transfer learning.	
	In~\cite{fu_energy_2021}, the authors consider a UAV-assisted IoT data collection system where a ground charging station wirelessly charges a UAV during the data collection task. They first formulate the problem as a Markov decision process (MDP), then propose a Q-learning algorithm to maximize the energy efficiency and system throughput. 	
%	The usage of Q-learning in~\cite{bouhamed_uav_2020} and~\cite{fu_energy_2021} makes their solution unscalable, especially for IoT systems with thousand devices. 

In~\cite{zhang_hierarchical_2021}, the authors aim to minimize the total working time of UAVs in a data collection system where the UAVs are used to collect data from backscatter sensor nodes. 
	If the UAVs' remaining energy is insufficient to complete the task, they can return to a charging station for charging. 	
	They first use the Gaussian mixture model to group IoT nodes into different clusters and formulate the trajectory optimization problem as a semi-Markov decision process. Then, to find the optimal policy for the UAV, deep reinforcement learning approaches are proposed. 
	In~\cite{xu_blockchain_2021}, the authors propose a blockchain-enabled UAV-assisted framework to provide security for IoT data collection network. A charging coin is introduced to reward UAVs when they successfully collect IoT data. Then, the UAVs can use collected coins to recharge their batteries at a charging station. 
	In addition, they develop an adaptive linear prediction model to reduce the number of transactions in the system, resulting in a decrease in energy consumption.	
	
All of the above works assume that the UAVs fly at a constant speed during the data collection process. 
	However, in practice, a UAV can choose different speeds during its data collection process depending on its surrounding environment. 
	Alternatively, the UAV's speed can strongly influence the system's efficiency because it has a substantial impact on energy consumption during the data collection process~\cite{shan_looking_2020}. 
	Thus, optimally controlling UAVs' speed can significantly improve the energy usage and data collection efficiency of the system, especially in UAV-assisted IoT data collection networks where UAVs have limited battery capacities.
	Unfortunately, this important factor is not investigated in all the above studies.
\subsubsection{Speed Control for UAVs}
In the literature, only a few works investigate the speed control problem for UAV-assisted IoT data collection networks~\cite{Gong2018JSAC, Pan2018Sensor, Lin2019JIoT}. 
	In~\cite{Gong2018JSAC}, the authors aim to minimize the flight time for a data collection task by jointly optimizing the UAV's speed, data collection duration, and the IoT devices' transmit power. 
	For a simple case where only one IoT node presents, they propose a water-filling policy and a bisection search to control the IoT node's power transmit power and the UAV's speed.
	Then, dynamic programming is adopted to address a general case with multiple IoT nodes.
	Their numerical results show that the UAV's optimal speed depends on the distance between sensors, sensors' energy, and the data upload requirements.
	In~\cite{Pan2018Sensor} and \cite{Lin2019JIoT}, the authors aim to maximize the data collection efficiency by controlling UAV's speed according to the IoT device density.
	In particular, the authors in~\cite{Pan2018Sensor} first introduce an analytical model for the transmission between the UAV and IoT nodes, then the UAV's speed is optimized based on this model. 
	In~\cite{Lin2019JIoT}, the authors reveal a tradeoff between system throughput and IoT devices' energy efficiency. 
	By optimizing the UAV's speed, altitude, as well as the MAC layer frame length, we can achieve the balance between the two conflict factors.
	
	All the above works (i.e.,~\cite{Gong2018JSAC,Pan2018Sensor,Lin2019JIoT}) statically optimize the UAV's speed during the IoT data collection process. Therefore, their algorithms need to rerun if there is any change in the environment, leading to a high computation complexity.
	The work in~\cite{li_onboard_2019} addresses this issue by using an MDP framework to adjust the UAV's speed during its data collection process. Furthermore, the UAV can wirelessly charge the IoT devices while collecting their data.
	This work aims to minimize the data packet loss by selecting the best devices to be charged and interrogated, together with the optimal UAV's speed. 
	Their simulation results show that the UAV's speed is proportional to the number of IoT devices and inversely proportional to the data queue lengths of IoT devices.		
	Similar to~\cite{Gong2018JSAC, Pan2018Sensor, Lin2019JIoT}, the study in~\cite{li_onboard_2019} does not consider the impacts of UAV's energy consumption and energy replenishment processes during the data collection task.
	It is worth highlighting that the energy replenishment process is a critical factor that cannot be ignored since the UAVs' energy is limited. 	
	
It can be observed that all of the aforementioned works	do not jointly optimize energy replenishment and speed control activities simultaneously.
	However, they are among the most important factors to achieve high efficiency in terms of energy and data collection for UAV-assisted IoT data collection networks. 
	In addition, regarding related works using reinforcement learning approaches to find optimal operation policies for UAV collectors, only the work in~\cite{bouhamed_uav_2020} proposes to use a transfer learning model to speed up the learning process. 
	However, impacts of transfer learning algorithms are not well investigated.
	Note that transfer learning does not always improve or even can cause negative impacts on the learning process~\cite{taylor_transfer_2009}.
	Furthermore, in~\cite{bouhamed_uav_2020}, they do not consider the UAV's speed control, one of the most important factors influencing to decisions of energy replenishment. 
	To fill this gap, the purpose of this paper is to provide a highly efficient solution based on Deep Reinforcement Transfer Learning for UAV-assisted IoT data collection networks. Our proposed approach can simultaneously optimize the UAV's speed and energy replenishment processes and allow the learned knowledge to be effectively ``shared'' and ``transferred''  between UAVs.
	  
\subsection{Contributions}
In this work, we aim to develop a novel framework that can not only address all the above problems but also allow to effectively implement on different UAVs in dissimilar environments. 
	In particular, in order to jointly optimize the speed control and battery replacement activities for a considered UAV under the dynamic and uncertainty of IoT data collection process, we propose a dynamic decision solution leveraging the Markov decision process (MDP) framework. 
	This framework allows the UAV to make optimal decisions (regarding the flying speed and battery replacement activities) based on its current observations about the surrounding environment. 
	Then, we adopt the Q-learning algorithm to find the UAV's optimal operation policy. 
	Although Q-learning can guarantee the convergence of the learning process, its convergence rate is slow, especially 
	in a highly complex problem as the one considered in this paper, i.e., we need to jointly optimize the speed and energy replenishment activities for the UAV. 
%	Specifically, it is originated from not only the high dimensional state and action spaces but also the complexity in making decisions to maximize the system performance, i.e., whether it should return for energy replenishment or continue collection data and how fast it should fly to collect IoT data.   
	In addition, Q-learning algorithms usually suffer from overestimation problems when estimating action values, especially for complicated problems with hybrid actions like what we consider in this paper (i.e., speed selection and energy replenishment actions)~\cite{Hasselt2016}. 
	Thus, we develop a highly-effective Deep Dueling Double Q-learning (D3QL) to address these challenges. The key ideas of D3QL are to (1) separately and simultaneously estimate the state values and action advantages, making the learning process more stable~\cite{Wang2016Dueling}, and (2) address the overestimation by using two estimators (e.g., deep neural networks), resulting in the stability of estimating action values.
	To further reduce the learning time and enhance the learning quality, we develop transfer learning techniques to allow the UAV to learn more knowledge from other UAVs learning in similar environments. 
	In addition, these techniques also help the UAV leverage knowledge obtained from different environments to improve its policy, making our solution more applicable and scalable. 
	Extensive simulation results demonstrate that our proposed solution, i.e., D3QL with transfer learning (D3QL-TL), can simultaneously optimize the energy usage and data collection, and thereby leading to the best performance compared to other methods. 
	To the best of our knowledge, this is the first study investigating a UAV operation control approach taking the dynamic of the IoT data collection, energy limitation, and impact of the energy replenishment process into considerations. 
	Our major contributions can be summarized as follows.
\begin{itemize}
	\item Propose a novel framework that allows the UAV to jointly optimize its flying speed and battery replacement activities. 
		In addition, this framework can not only allow the UAV to dynamically and automatically make optimal decisions through real-time interactions with the surrounding environment but also enable the ``share/transfer'' learning knowledge among UAVs working in the same and/or similar environments. 
	\item Develop an MDP framework to overcome the dynamic and uncertainty of data collection and energy replenishment processes, and introduce a reinforcement learning algorithm to assist the UAV in finding the optimal operation policy without requiring complete knowledge of its surrounding environment. 
	\item Develop a highly-effective deep reinforcement learning algorithm leveraging recent advances of deep double Q-learning and dueling neural network architecture to stabilize the learning phase, thereby quickly obtaining an optimal operation policy for the UAV.
	\item Develop advanced transfer learning techniques that allow UAVs to ``share'' and ``transfer'' their learning knowledge, thereby reducing the learning time and improving learning quality for the UAV.
		Furthermore, these techniques help UAVs to utilize the knowledge and information learned from different environments, thereby making our approach more scalable and applicable in practice. 
	\item Perform extensive simulations to demonstrate the efficiency of our proposed approaches and reveal critical elements that can significantly impact on the performance of UAV-assisted IoT data collection networks.
\end{itemize}

	The rest of this paper is structured as follows. The system model and operation control formulation are described in Section~\ref{sec:sysmodel} and~\ref{sec:markov}, respectively. Section~\ref{sec:LA} presents the proposed Q-learning, D3QL, and D3QL-TL algorithms. Then, the simulation results are analyzed in Section~\ref{sec:PE}. Finally, Section~\ref{sec:conclusions} concludes our paper.

%%%%%%%%%%%%%%%%%%%%%%%%%%%%%%%%%%%%%%%%%%%%%%%%%
%%%%%%%%%%%%%%%%%%%%%%%%%%%%%%%%%%%%%%%%%%%%%%%%%

\section{System Model}
\label{sec:sysmodel}
\begin{figure}[t]
	%	\captionsetup{singlelinecheck=off}
	\centering
	\includegraphics[width=0.9\linewidth]{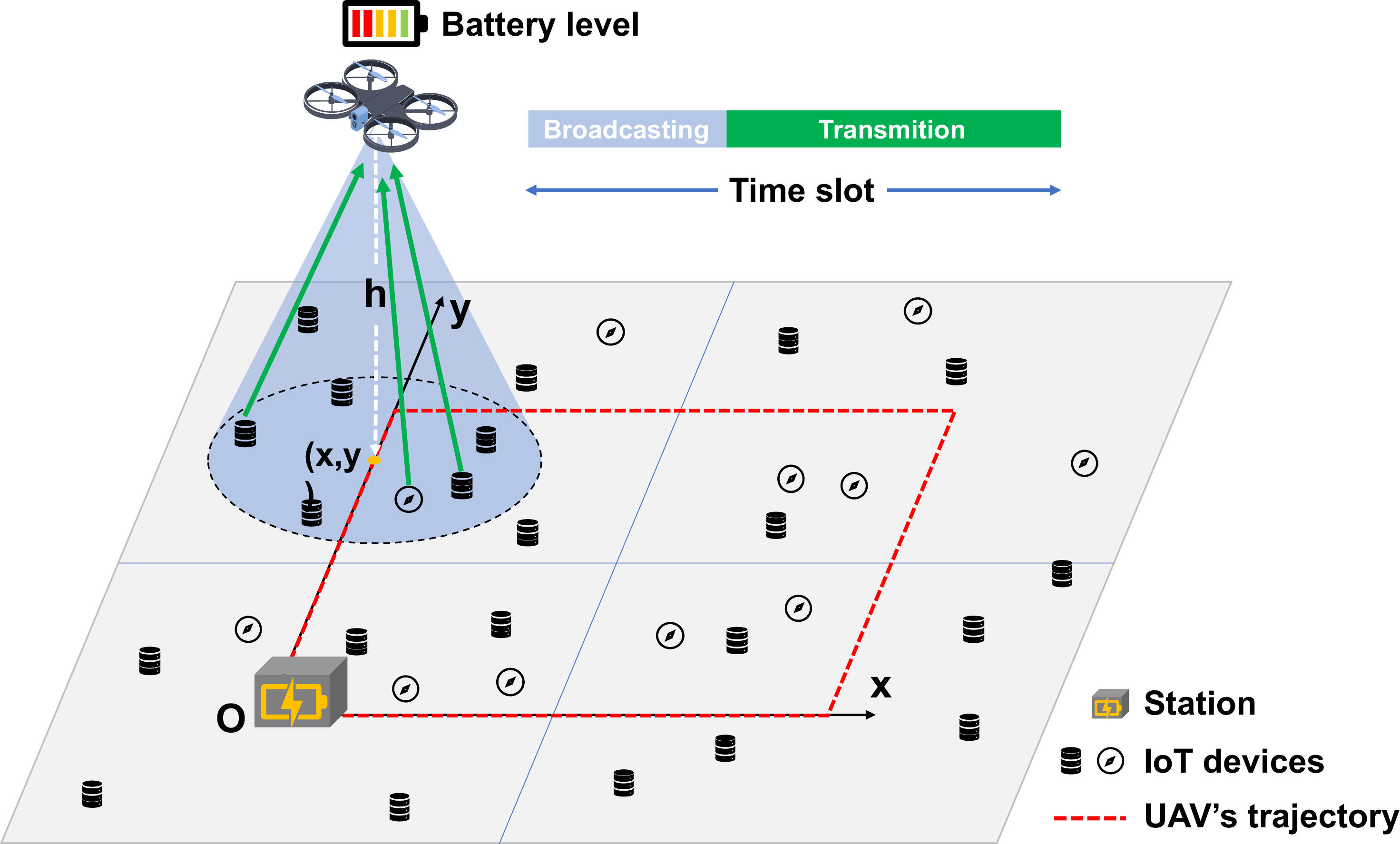}
	\caption{System model for UAV-assisted IoT data collection network.}
	\label{fig.system_model}
\end{figure}
%%%%%%%%%%%%%%%%%%%%%%%%%%%%%%%%%%%%%%%%%%%%%%%%	
%%%%%%%%%%%%%%%%%%%%%%%%%%%%%%%%%%%%%%%%%%%%%%%%	
%\subsection{System Model}
In this work, we consider a UAV-assisted IoT data collection system where a UAV is deployed to collect IoT data over a considered area, as illustrated in Fig.~\ref{fig.system_model}.
	We assume that the considered area is divided into $N$ zones.
	The IoT devices are distributed randomly over these zones to execute various tasks, e.g., sensing temperature and humidity. 	
	In practice, the numbers of IoT nodes in these zones are different due to different sensing demands in these zones. 
	We consider time to be slotted (as in~\cite{Pan2018Sensor, Lin2019JIoT}) with an equal duration, and each time slot is split into two consecutive intervals, broadcasting and transmission, respectively. 
	In the broadcasting interval, the UAV uses a dedicated channel to broadcast a wake-up signal~\cite{khodr_energy_2017} to all IoT nodes in its communication range. 
	After acquiring this signal, these nodes will send their data to the UAV during the transmission interval. 
	We assume that the communication link from IoT devices to the UAV adopts the OFDMA technique, while the communication link from the UAV to IoT devices uses the OFDM technique, as in \cite{Wu2018OFDMA}. 
	In this way, the IoT devices can simultaneously transmit data to the UAV. 
	Let $p_n$ denote the probability of a data packet successfully collected by the UAV in a time slot in zone $n$. 
	Because the IoT nodes are distributed unevenly over $N$ zones, $p_n $ may vary over these zones. 

As considering a UAV-assisted IoT data collection network, we can assume that the UAV flies at a fixed altitude $h$,  imposing by a regulation for safety considerations (similar as that of~\cite{Liu2020Distributed, elmagid_deep_2019, Wu2018OFDMA, Pan2018Sensor, Gong2018JSAC}) and follows a predefined trajectory (as in \cite{Gong2018JSAC, Pan2018Sensor, Lin2019JIoT}) to sweep through all IoT devices of the system in each round. 
	In this paper, unlike \cite{Pan2018Sensor, Lin2019JIoT, Gong2018JSAC}, we consider a more realistic scenario where the UAV is equipped with a battery that has limited energy storage. 
	It is worth mentioning that as the energy consumption for wireless data collection process (i.e., broadcasting the wake-up signal and receiving data packets) is much lower than that of the flying operation~\cite{shan_looking_2020}, we only focus on optimizing energy consumption for the UAV's flying operation in this work.
	In a time slot, we assume that the UAV's velocity is constant (as in \cite{Wu2018OFDMA,Gong2018JSAC}),  
	but in different time slots the UAV can choose to fly at different speeds, e.g., $v_{w} = \{v_1, \dots, v_A\}$.
	Each speed may cost a different amount of energy.
	For example, if the UAV flies faster, it may use more energy per time slot~\cite{shan_looking_2020}.

When the UAV's energy is depleted, it will fly back to the charging station placed at a fixed location to replace the battery, as illustrated in Fig.~\ref{fig.system_model}.
	Furthermore, during the flight, the UAV can decide to go back to the charging station to change the battery, for example, when it is near the charging station and its energy level is low.	
	Once the energy replenishment process is accomplished, the UAV will fly back to its trajectory and continue its task. 
	Here, we consider that the UAV has a maximum of $E$ energy units for its operation (i.e., flying to collect IoT data) and a backup energy storage for flying back to the charging station for battery replacement. 	
	During the energy replenishment process, including back-and-forth flights and battery replacement, the UAV cannot collect data. 
	Suppose that it takes the UAV $t_f$ and $t_{b}$ time slots to fly from its current location to the station and to replace the battery, respectively.
	Indeed, the battery replacement time, i.e., $t_b$, may be known in advance, while the return flight time, i.e., $t_f$, is highly dynamic depending on the distance between UAV's current location and the charging station.
	In addition, $t_f$ also depends on the return speed of the UAV, denoted by $v_r$.
	Assuming that UAV flies with a constant speed when returning to the station, the duration of energy replenishment is calculated by the equation $t_{e} = 2t_f + t_b$.
	Therefore, the energy replenishment process is also dynamic due to the dynamic of flying time $t_f$. 
%		It is worth highlighting that our proposed algorithms help the UAV learn the optimal policy without prior information about $\mathbf{p}$, $t_f$, and $t_r$. 
%		The UAV will gradually learn this knowledge through its interactions with the environment.

In practice, the surrounding environment is highly dynamic and uncertain.
	Specifically, the UAV does not know the probabilities of receiving a packet in different areas in advance, as they are very uncertain depending on sensing tasks. 
	It is important to note that the UAV may collect more data when moving in a zone with a high probability of receiving packets, i.e., a high value of $p_n$.
	As a result, to maximize the data collection efficiency, the UAV must gradually learn this knowledge in order to adapt its operations accordingly, e.g., flying speed and energy level status.	
	Moreover, the returning flight time depends on the distance between the UAV's current position and the station, which is highly dynamic. 	  
	Therefore, if the UAV appropriately decides when to return for battery replacement (e.g.,  when it is near the station and its energy level is low), the energy replenishment time will be reduced significantly, resulting in high system performance. 
	In contrast, if the UAV goes back to replace its battery when its energy level is high and it is far from the station, it will waste both time and energy, leading to low system performance.  
	Thus, optimizing the UAV's operations to maximize the long-term system performance is a very challenging task.
	In the following sections, we will present our proposed learning algorithms that can effectively and quickly obtain the UAV's optimal operation policy under the limited energy of the UAV and the uncertainty of the data collection process. 

%--------------------------------------------------------------------------------------------------------------------------
%--------------------------------------------------------------------------------------------------------------------------
\section{Optimal Operation Control Formulation}
\label{sec:markov}
To overcome the uncertainty and highly dynamic of the data collection and energy replenishment processes under the limited energy storage of the UAV, we formulate the UAV's operation control problem as the Markov decision process (MDP) framework. 
	Typically, an MDP is determined by three components, state space $\mathcal{S}$, action space $\mathcal{A}$, and immediate reward function $r$.
	Based on the MDP framework, at each time slot the UAV can dynamically make the best actions (e.g., flying at appropriate speeds or return for battery replacement) based on its current observations (i.e., its location and energy level) to maximize its long-term average reward without requiring complete information about data collection and energy replenishment processes in advance.
%%%%%%%%%%%%%%%%%%%%%%%%%%%%%%%%%%%%%%%%%%%%%%%%	
%%%%%%%%%%%%%%%%%%%%%%%%%%%%%%%%%%%%%%%%%%%%%%%%
\subsection{State Space}
\label{subsec:state_space}
In this work, we aim to maximize the efficiency of collecting data and energy usage efficiency, and thus there are some important factors which we need to take into considerations. 
	The first important factor is the current location of the UAV. 
	The main reason is that the UAV's location can reveal important information about the expected amount of data that can be collected by the UAV and the time it takes if the UAV chooses to fly back to the station for battery replacement.
	Specifically, the UAV will likely collect more data when moving in a zone with a high probability of receiving data than in a zone with a low probability of receiving data.
	In addition, the farther the distance between the UAV and the station is, the more time it takes to travel between these two positions.		
	As mentioned above, the UAV always flies at a fixed altitude so that the UAV's position can be given by its 2D projection on the ground, i.e., $(x,y)$ coordinates.
	The second crucial factor is the current UAV's energy level, denoted by $e$, which affects the decision of the UAV at every time slot. 
	For example, the UAV should not select the battery replacement action (i.e., return the station to replace the battery) unless its energy level is low.
	Otherwise, it will waste time and energy for the flying back trip. 	
To that end, this information is embedded into the state space of the UAV, which can be defined as follow:
	\begin{equation}
		\begin{aligned}
			\mathcal{S} = \Big\{ (x,y,e): x \in \{0,\ldots, X\}; y \in \{0,\ldots, Y \}; \\ \text{ and } e \in \{0,\ldots, E \}  \Big\}\cup \{(-1,-1,-1)\},
		\end{aligned}
	\end{equation}
	where $X$ and $Y$ are the maximum x and y coordinates of the UAV's trajectory, and $E$ is the maximum energy capacity of the UAV.
	As a result, the system state can be indicated by a tuple $s = (x,y,e) \in \mathcal{S}$.
	Moreover, because of the energy replenishment process, it is necessary to introduce a special state, i.e., $s=(-1,-1,-1)$. 
	This special state is only visited when UAV's energy is depleted (i.e., UAV's state is $(x,y,0)$) or if the UAV selects the battery replacement action.
	Then, after the energy replenishment process completes, the UAV will return to the previous position (where it decided to go back for battery replacement or where its energy is depleted) with a full battery, i.e., $s=(x,y,E)$. 
	This design ensures that the system process is continuous, i.e., no terminal state.
%=========================================
%=========================================
\subsection{Action Space}
\label{subsec:action_space}
During the operation, to maximize the system performance in terms of energy usage and data collection efficiency, the UAV needs to not only choose the most suitable flying speed but also decide when to go back to the station to replace the battery. 
	It is worth mentioning that given different states, the action spaces for these states may be different.
	For example, at a non-working state, i.e., $s=(-1,-1,-1)$, the UAV cannot select a flying speed.
	Instead, the possible action at this state is to stay ``idle'' until the UAV returns to its trajectory with a full battery.
	In other words, the UAV will stay at the non-working state after performing an ``idle'' action until the energy replenishment process completes.
	As a result, we can define the action space for the UAV as follows: 
	\begin{equation}
		\mathcal{A} \triangleq \{a: a \in \{-1, 0,1, \ldots, A\}\},
	\end{equation}
	where action $a=-1$ is to indicate the ``idle'' action and action $a=0$ is to express that the UAV will choose to return to the station for replacing the battery, namely battery replacement action.
	Actions $a=\{1,\ldots, A\}$ are to represent the speed level that the UAV selects to fly at the current time slot.
	In addition, given the state $s\in \mathcal{S}$ the action space based on state $s$, i.e., $\mathcal{A}_S$, consists of all possible actions that are feasible at this state.
	Thus, we can express $\mathcal{A}_S$ as follows:
	\begin{equation}
		\mathcal{A}_S = \left\{\,\begin{array}{ll}
			\{-1\}, 				&\mbox{if $s=(-1,-1,-1)$,} \\
			\{0, \ldots, A\}, 	&\mbox{otherwise.}
		\end{array}	\right .
	\end{equation}
	%=========================================
%=========================================
\subsection{Reward Function}
\label{subsec:reward_function}
As discussed above, two main actions (i.e., flying speed and battery replacement actions) have significant effects on the system performance.
	Specifically, choosing an appropriate flying speed at each time slot can maximize the efficiency of the collecting data process as well as energy usage.
	Alternatively, selecting the right time to return for battery replacement can reduce the energy replenishment time, thereby improving the overall system performance. 
	For example, when the UAV is flying near the charging station and its energy is low, it should return to the charging station for battery replacement. 
	Therefore, our proposed immediate reward function consists of (1) speed selection reward function, i.e., $r^a_t$, and (2) battery replacement reward function, i.e., $r^b_t$, as follows:
	\begin{align}
		\label{eq:total_reward_function}
		&r_t(s_t, a_t) = \left\{\,	\begin{array}{ll} 
			r^{a}_t(s_t, a_t)  ,				&	\mbox{if $a_t \in \{\mathcal{A}\setminus\{0\}\}$,}	\\
			r^{b}_t(s_t, a_t),	&	\textrm{if $a_t=0$,}\\
			0,					& 	\mbox{otherwise,}
		\end{array}	\right.
	\end{align}
	where $a_t$ is the selected action at time $t$.

	\subsubsection{Speed Selection Reward Function}
	Since our objective is to maximize the system performance by striking a balance between the data collection efficiency and energy usage efficiency, the speed selection reward function needs to capture this information.
		We define data collection efficiency as the number of collected data packets over a time slot and operation status of the UAV.
		For example, given a time slot, if the UAV is moving to collect data, it will receive a working reward, denoted by $\Omega>0$.
		Otherwise, it will not receive the working reward, i.e., $\Omega=0$. 		
		In this way, the working reward will encourage the UAV to collect data rather than return and then wait at the station for the battery replacement. 
		The energy usage efficiency can be represented by the cost of choosing a flying speed, i.e., energy consumed by the UAV to fly at speed $a$ during a time slot $t$. 
		Clearly, the selected speed determines the energy consumption per time slot of the UAV, e.g., a low speed will cost the UAV less energy per time slot than that of a high speed~\cite{shan_looking_2020}. 
		At time slot $t$, the cost of performing action $a$ is denoted by $m^a_t$, and the number of collected data packets at the current state $s$ is denoted by $d^s_t$.
		Thus, the speed selection reward function can be expressed by:  		
		\begin{align}
			\label{eq:speed_reward_function}
			&r^a_t(s_t,a_t)= \Omega +	w_1 d^s_t - w_ 2 m^a_t, \quad \mbox{if $a_t \in \mathcal{A}\setminus\{0\}$},
		\end{align}	
		where $w_1$ and $w_2$ are the weights to balancing between collected data and energy consumption of the UAV.
		It is worth noting that these weights can be defined in advance based on the service provider's requirements.
		For example, in case if the data is more important and valuable than energy, we can set the value of $w_1$ to be higher than that of $w_2$. 
		In contrast, if the energy is scarce, we can set the value of $w_1$ to be lower than that of $w_2$.
		Therefore, the speed selection reward function can capture the UAV's data collection efficiency and energy consumption efficiency.

	\subsubsection{Battery Replacement Reward Function}
	Although the speed selection reward function gives the UAV sufficient information to learn the optimal speed control, it cannot help the UAV to learn when it is good to return to the station for the battery replacement.
		First, if the UAV performs the battery replacement action, the values of $\Omega$ and $d^s_t$ in~\eqref{eq:speed_reward_function} will be zero, leading to a negative value of $r^a_t$. 
		Consequently, the UAV may consider this action as a bad choice and will not choose it in the future.
		Second, the speed selection reward function is unable to guide the UAV to learn where is good to return for replacing its battery, e.g., the further from the station the UAV is, the smaller reward for battery replacement action it may receive. 
		Therefore, battery replacement action needs a different reward function, which needs to take into account of both the UAV's current energy level, i.e., $e$, and the distance between its current position and the station, i.e., $l$.  
		\begin{figure}[t]
			\centering
			\includegraphics[width=0.8\linewidth]{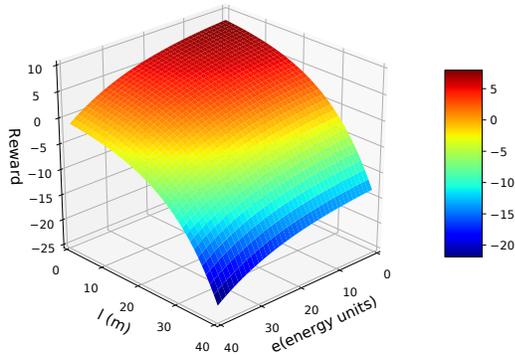}
			\caption{An example of the proposed battery replacement reward function.}
			\label{fig:b_reward_function}
		\end{figure}
		However, in practice, it is challenging to design such a reward function for battery replacement action because the complex relationship between $e$ and $l$ makes more difficulties for the UAV to decide whether it should go back or keep flying to collect data.
			In particular, the UAV may choose the battery replacement action if both $e$ and $l$ are small, while the UAV should continue its collection task if any of these factors is large.
			To that end, in the following, we propose a reward function for battery replacement action that can address this problem, and to the best of our knowledge, this is the first work in the literature addressing the battery replacement problem for UAV-assisted IoT data collection networks.
			
		Suppose the UAV decides to return for battery replacement at time $t$, at which its energy and the distance to the station are $e_t$ and $l_t$, respectively. 
			Then, its immediate reward is derived from the battery replacement reward function as follows:
			\begin{align}
				\label{eq:return_reward_function}
				&r^{b}_t(s_t,a_t) = c - w_3\exp(w_4e_t) - w_5\exp(w_6l_t), \quad \text{if } a_t=0,
			\end{align}
			where $c$ is a constant controlling the maximum value of $r^{b}_t$, which may affect the learning policy of the UAV. 
			For example, if $c$ is smaller than the smallest value of $r^a_t$, the value of returning reward function is always lower than that of the speed selection reward function, making the return action always ``worse'' than those of the speed selection actions. 
			The second and third terms in~\eqref{eq:return_reward_function} express the influence of the energy level $e_t$ and the distance $l_t$ to the UAV's decision for battery replacement.
			The tradeoff between the energy level $e_t$ and the current distance $l_t$ is controlled by four weights, i.e., $w_3,w_4,w_5,w_6$.  
			Fig.~\ref{fig:b_reward_function} demonstrates an example of the proposed battery replacement reward function in which $c=10$, $w_3=w_5=1$, $w_4=0.06$, and $w_6=0.08$.
			It can be observed that as $e$ and $l$ are large (e.g., greater than $36$ and $17$, respectively), the UAV will receive negative rewards, meaning that the UAV is not encouraged to return to replace its battery if its energy is high or it is far from the station.
			In this way, this function will encourage the UAV to return to the station for replacing the battery when its current energy and distance to the station are small. 
\begin{figure*}[t!]
	\centering
	$\begin{array}{cc}
		\includegraphics[width=0.4\linewidth]{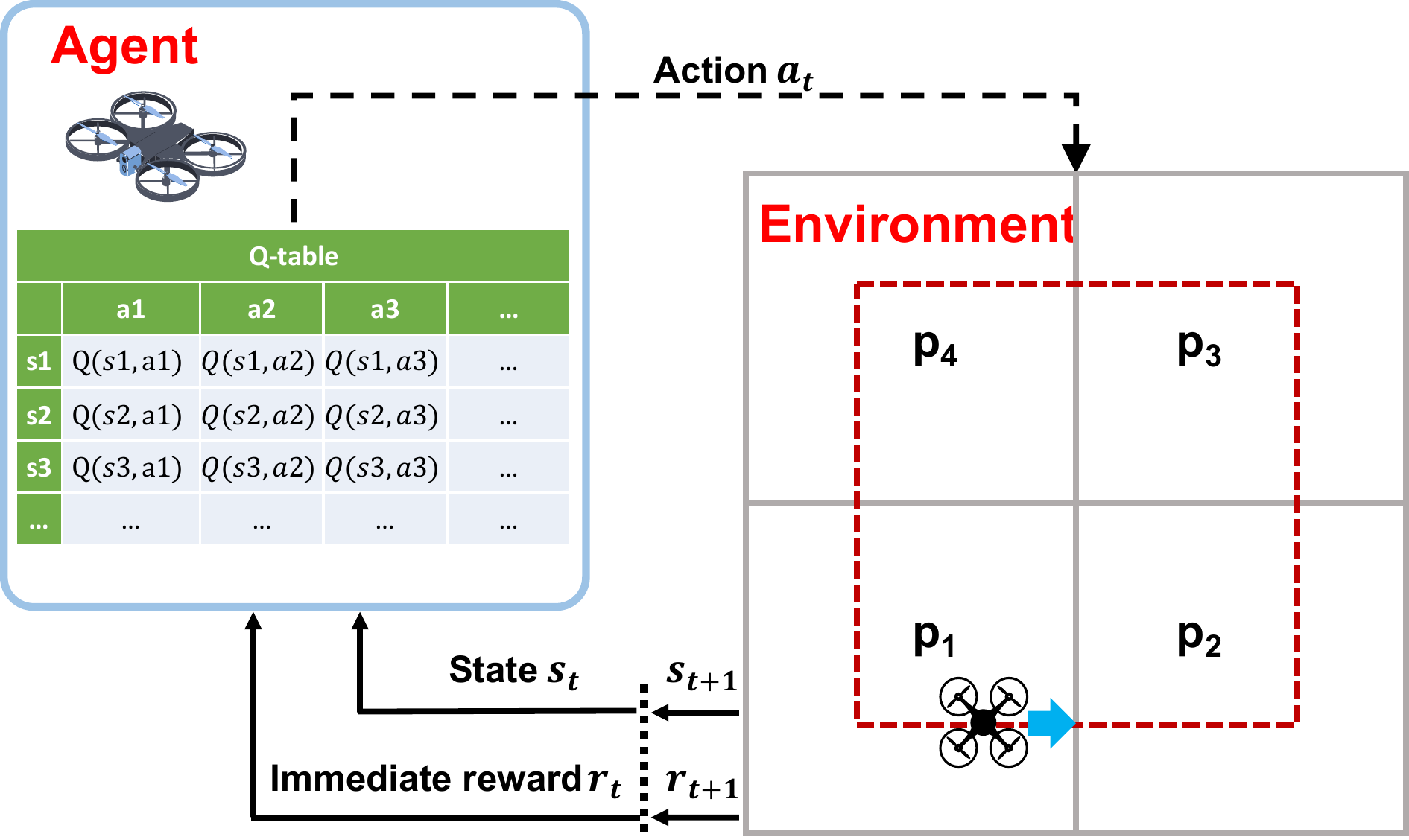}		 & \qquad
		\includegraphics[width=0.4\linewidth]{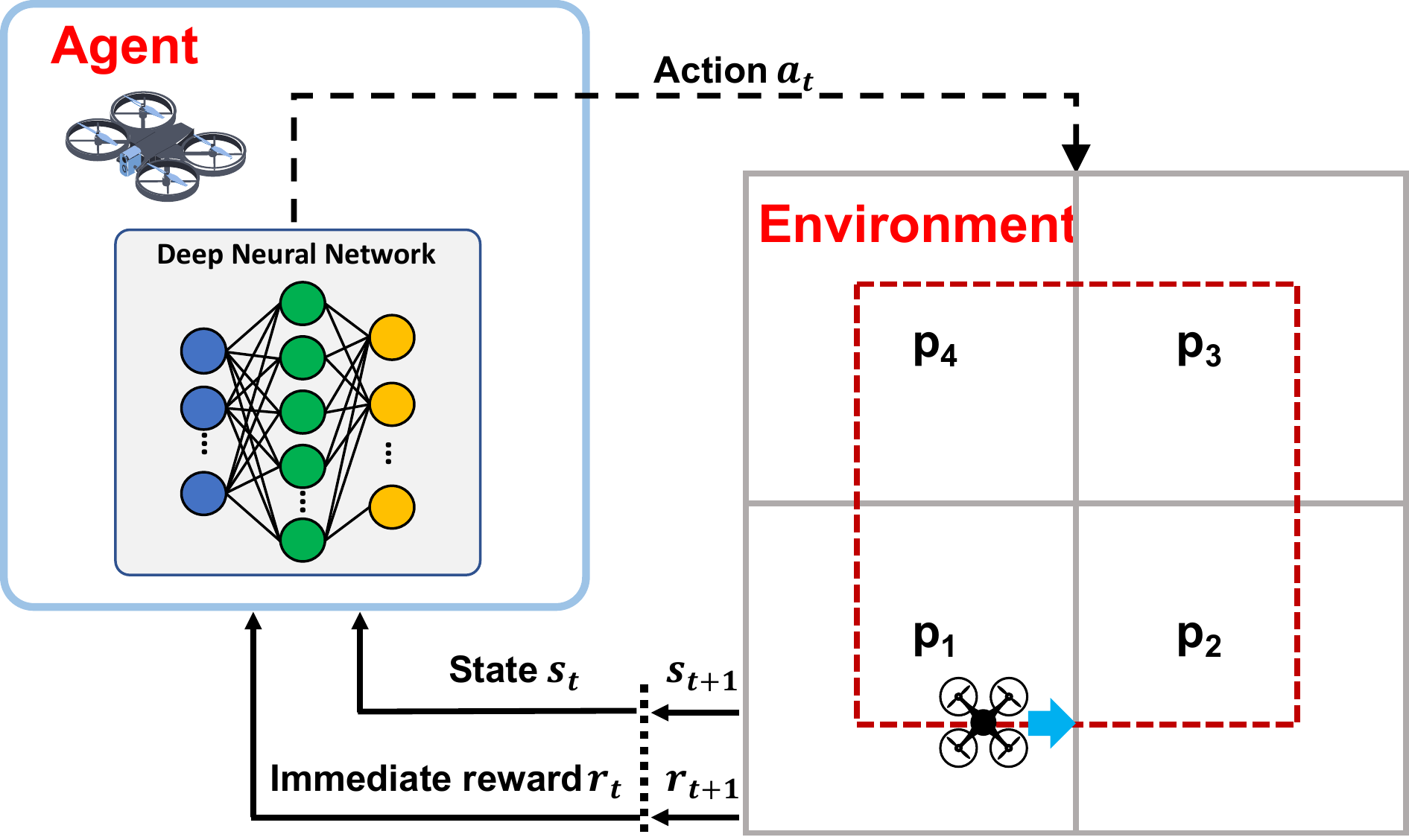}		\\
		\text{(a) Q-learning}	& \qquad
		\text{(b) Deep Q-learning}
	\end{array}$
	\caption{(a) Q-learning and (b) Deep Q-learning based models.}
	\label{fig.Qlearning_DQN}
\end{figure*}			
\subsection{Optimization Formulation}
	In this paper, our aim is to maximize the average long-term reward function by finding a UAV's optimal policy $\pi^*$, i.e.,~\mbox{$\pi^*:\mathcal{S} \rightarrow \mathcal{A} $}. In particular, given the UAV's current energy and location, $\pi^*$ determines an action that maximizes the long-term average reward function as follows:
	\begin{eqnarray} 
		\label{eq:average_reward}
		\max_\pi \quad	{\mathcal{R}}(\pi)	=	\lim_{T \rightarrow \infty} \frac{1}{T} \sum_{t=1}^{T} {\mathbb{E}} \big( r_t(s_t,\pi(s_t)) \big),	\label{eq:cmdp_obj_2}
	\end{eqnarray}
	where ${\mathcal{R}}(\pi)$ is the long-term average reward obtained by the UAV according to the policy $\pi$, $\pi(s_t)$ is the selected action at state $s$ at time slot $t$ based on policy $\pi$, and $r_t(s_t,\pi(s_t))$ is the immediate reward by following policy $\pi$ at time slot $t$. 
	Thus, the optimal policy $\pi^*$ will assist the UAV in dynamically making the optimal action according to its
	current observation, i.e., its position and energy level.

%	Obviously, the MDP with the state space $\mathcal{S}$ defined in Section.~\ref{subsec:state_space} is irreducible, i.e., the process can go from a given state to any other states after $M$ steps. 
%	Therefore, with any policy $\pi$, the long-term average reward ${\mathcal{R}}(\pi)$ is well-defined and does not rely on the beginning state.
%--------------------------------------------------------------------------------------------------------------------------
%--------------------------------------------------------------------------------------------------------------------------	
\section{Optimal Operation Policy for UAVs with Reinforcement Learning Approaches}
\label{sec:LA}
In the considered problem, the UAV does not have complete information about the surrounding environment in advance, such as the data arrival probabilities and the charging process, to obtain the optimal policy. 
	To deal with this challenge, we develop reinforcement learning-based algorithms that can help the UAV gradually learn the optimal policy without requiring complete information about surrounding environments in advance. 
	In particular, we first present a background on reinforcement learning with a conventional Q-learning algorithm. Then, we describe the proposed D3QL in details. Finally, we discuss our proposed transfer learning techniques.

	\subsection{Reinforcement Learning}	
		Reinforcement learning (RL) is widely used to address MDP problems. 
			In particular, RL-based algorithms can assist the UAV in acquiring the optimal policy through interactions with the environment, e.g., the number of data packets collected from different locations. 
			Suppose that from state $s \in \mathcal{S}$, the UAV follows policy $\pi$, the value of this state is then determined by the state-value function for policy $\pi$ as follows: 
			\begin{equation}
				\label{eq:V-function}
				\mathcal{V}^\pi(s) = \mathbb{E}_\pi \Bigg[\sum_{t=0}^{\infty}\zeta^t r_t(s_t,a_t) |s_0 = s, s \in \mathcal{S} \Bigg],
			\end{equation}
			where $\mathbb{E}$ is an expectation function with respect to policy $\pi$, $r_t$ is the immediate reward received after performing action $a_t$ at state $s_t$, and $\zeta \in [0,1]$ is the discount weight that specifies the trade-off between the importance of immediate reward and that of the future rewards ~\cite{Watkins1992QLearning}. 
%			For example, the immediate reward is more important than future rewards if $\zeta$ is close to $0$, while the future rewards become more important when $\zeta$ approaches $1$.
			While the state-value function specifies how good to be of a state, the action-value function decides the value of performing an action at a state as follows:
			\begin{equation}
				\label{eq:Q-function}
				\mathcal{Q}^\pi(s,a) = \mathbb{E}_\pi \Bigg[\sum_{t=0}^{\infty}\zeta^t r_t(s_t,a_t) |s_0 = s, a_0 = a \Bigg].
			\end{equation}
			Specifically, the action-value function is the expected sum of discounted future rewards obtained by starting from state $s$, performing action $a$ according to policy $\pi$.	
			Thus, the relationship between state-value and action-value functions can be expressed as follows:
			\begin{equation}
				\label{eq:relationship_Q_V-function}
				\mathcal{V}^\pi(s) = \sum_{a \in \mathcal{A}} \pi(a|s)\mathcal{Q}^\pi(s,a), \quad \forall s \in \mathcal{S},
			\end{equation}
			where $\pi(a|s)$ is the probability of selecting action $a$ at state $s$ under policy $\pi$.
			The purpose of RL is to find an optimal policy $\pi^*$ that maximizes a long-term cumulative reward, e.g., the long-term average reward in \eqref{eq:average_reward}. 
			In practice, we can obtain more than one optimal policy for a given MDP. 
			However, they all have the same state-value and action-value functions, namely optimal state-value function $\mathcal{V}^*(s)$ and optimal action-value function $\mathcal{Q}^*(s,a)$,~\cite{Sutton1998Reinforcement}, i.e.,
			\begin{align}
				\mathcal{V}^*(s) &\triangleq \max_\pi \mathcal{V}^\pi(s),	 \quad\quad~\forall s \in \mathcal{S},\\
				\mathcal{Q}^*(s,a) &\triangleq \max_\pi \mathcal{Q}^\pi(s,a),\quad~\forall s \in \mathcal{S}~\textrm{and}~ \forall a \in \mathcal{A}.
			\end{align}
			According to \cite{Sutton1998Reinforcement}, under the optimal policy, the optimal state-value
			function of a state equals to the optimal action-value of the best action at this state, i.e.,
			\begin{align}
				\mathcal{V}^*(s) &\triangleq \max_{a \in \mathcal{A}} \mathcal{Q}^*(s,a), \quad~\forall s \in \mathcal{S}.
			\end{align}
			Suppose that the optimal action-value function $\mathcal{Q}^*(s,a)$ is acquired, it is straightforward to obtain the optimal policy $\pi^*$ at state $s$ by taking an action that maximizes $\mathcal{Q}^*(s,a)$.
			In the following, to help the UAV obtain an optimal policy, we present our proposed algorithms that can learn optimal Q-values through interactions with surrounding environments.  

	\subsection{Q-learning for the UAV Dynamic Operation Control}
	\label{subsec:Q-learning}
		In RL, Q-learning is one of the most widely used algorithms because it can guarantee to converge to the optimal policy after the learning process~\cite{Watkins1992QLearning}.
		This algorithm aims to learn an action-value function, named Q-function and denoted by $\mathcal{Q}$, that directly approximates the optimal action-value function, i.e., $\mathcal{Q}^*(s, a)$, regardless of the policy that the UAV is following.  
		Therefore, Q-learning is an off-policy method.
		The details of Q-learning are provided in Algorithm~\ref{alg:qlearning}.
		In particular, Q-learning represents Q-function by a table, called Q-table.
		Each Q-table's cell stores an estimated value of Q-function, called Q-value, for one state-action pair, i.e., $\mathcal{Q}(s, a)$, as illustrated in Fig.~\ref{fig.Qlearning_DQN}(a). 
		These Q-values are updated through interactions with the environment. 
		Specifically, suppose that at time step $t$, the UAV selects action $a_t$ base on $\epsilon$-greedy policy.
%		, in which the value of $\epsilon$ trade-offs between the exploration, i.e., selecting a random action, and the exploitation, i.e., selecting the best-known action base on the Q-value estimates. 
%		For example, the larger the $\epsilon$ is, the higher randomness in selecting action is.
%		At the first step, $\epsilon$ is set at $1$, equivalent to randomly selecting an action, and then it gradually decreases to a small value, e.g., $0.01$.				
		After executing action $a_t$, the UAV observes next state $s_{t+1}$ and immediate reward $r_t$.
		Then, the Q-value for this state-action pair is updated by:	
		\begin{equation}
			\label{eq:updateQfunction}
			\begin{aligned}
				\mathcal{Q}_{t}(s_t,a_t) \leftarrow &\mathcal{Q}_t(s_t,a_t) +\beta_t\big[r_t(s_t, a_t) \\&+ \zeta\max_{a_{t+1}} \mathcal{Q}_t(s_{t+1}, a_{t+1}) - \mathcal{Q}_t(s_t,a_t)\big],
			\end{aligned}
		\end{equation}
		where $\beta_t$ is the learning rate controlling how much the new information observed from the environment will be used to update the current estimated Q-value.
			Here, the new information is the difference between the current estimated Q-value (i.e., $\mathcal{Q}_t(s_t,a_t)$) and the target Q-value (i.e., $G_t=r_t(s_t, a_t) + \zeta\max_{a_{t+1}} \mathcal{Q}_t(s_{t+1}, a_{t+1})$), which is called temporal difference (TD).
		In \cite{Watkins1992QLearning}, $\mathcal{Q}(s,a)$ is proved to converge with probability 1 to $\mathcal{Q}^*(s,a)$ if it is iteratively updated by \eqref{eq:updateQfunction} and the learning rate $\beta_t$ satisfies~\eqref{eq:learning_rate_rules}.
		
		\begin{equation}
			\label{eq:learning_rate_rules}
			\beta_t \in [0,1), ~\sum_{t=1}^{\infty}\beta_t = \infty, \mbox{ and } \sum_{t=1}^{\infty} ( \beta_t  )^{2} < \infty.
		\end{equation}
		\begin{algorithm}[t]
		\caption{Q-learning for the UAV Dynamic Operation Control}
		\label{alg:qlearning}
		\begin{algorithmic}[1]
			\STATE Initialize discount factor $\zeta$, learning rate $\beta$, and $\epsilon$.
			\STATE Initialize Q-table's cells with an arbitrary values, e.g., zero for all cells.
			\FOR{\textit{t = 1 to T}}
			\STATE Observe current state $s_t$.
			\STATE Choose action according to $\epsilon$-greedy policy as follow:
				\begin{align}
				a_t = \left\{
					\begin{array}{ll}
						\mbox{random action }a\in\mathcal{A},&\mbox{with probability }\epsilon,\\
						\underset{a \in \mathcal{A}}{\text{argmax}}~\mathcal{Q}(s_t,a),&\mbox{otherwise}.						
					\end{array}	\right.
					\label{eq:epsilon_greedy}
				\end{align}
			\STATE Execute $a_t$, observe reward $r_t$ and next state $s_{t+1}$.
			\STATE Update $\mathcal{Q}(s_t,a_t)$ by~\eqref{eq:updateQfunction}.  
			\STATE Replace $s_t$ $\leftarrow$ $s_{t+1}$.
			\STATE Decrease $\epsilon$.
			\ENDFOR
		\end{algorithmic}
	\end{algorithm}
		Nevertheless, using a table to store Q-values for all state-action pairs makes the Q-learning algorithm facing the curse-of-dimensionality problem. 
		Specifically, Q-learning usually suffers from a long learning time or sometimes is even unable to achieve the optimal policy for the UAV because of the high dimensional state space  as in the considered problem.
		In addition, the uncertainties of data arrival and energy replenishment processes make more challenges for the UAV to achieve the optimal policy.
		Another issue of Q-learning is overestimations caused by the max operator when calculating the target Q-value $G_t$ \cite{Hasselt2016}.
		Specifically, since Q-value is an estimation, an action can get a higher Q-value than its true value. 
		If this action has the highest Q-value, it will be chosen to compute the target Q-value, leading to the overestimation when the Q-function is updated by \eqref{eq:updateQfunction}. 
		It may not be a big issue if the overestimations are uniform for all actions because all Q-values are shifted by the same amount. 
		However, in practical scenarios such as a UAV-assisted IoT data collection process, the overestimations are typically not uniform~\cite{Hasselt2016}, and thus it may significantly slow down the learning process of the UAV.		
		To overcome these challenges, we propose a highly-effective Deep Dueling Double Q-learning (D3QL) algorithm to quickly obtain the UAV's optimal operation policy, and thereby maximizing the system performance in terms of data collection and energy usage efficiencies.

	\subsection{Deep Dueling Double Q-learning}
		In this subsection, we propose a Deep Dueling Double Q-learning algorithm (D3QL) to address the slow convergence rate and overestimations in Q-learning. 
			In D3QL, we adopt three innovation techniques, including deep Q-learning~~\cite{Mnih2015Human}, dueling deep neural network architecture~\cite{Wang2016Dueling}, and double deep Q-learning~\cite{Hasselt2016}.
			The usage of Q-table to approximate Q-values makes the Q-learning algorithm struggling in finding the optimal policy if the state and action spaces are very large, especially under the scenario considered in this work.
			Therefore, deep neural networks (DNNs) have been introduced recently to address this problem, forming a new group of RL algorithms called deep reinforcement learning (DRL).
			In~\cite{Mnih2015Human}, the authors propose a deep Q-learning algorithm (DQL), in which the optimal state-action values~$\mathcal{Q}^*(s, a;\xi)$ are approximated by a DNN, called DQN, as shown in Fig.~\ref{fig.Qlearning_DQN}(b).			 
			In particular, feeding a state $s$ to DQN will return Q-values for all actions at this state, and each Q-value is given by one neuron of the output layer of DQN. 
			In addition, DQL employs a memory to store past experiences, i.e., sets of $<s_t, a_t, r_t, s_{t+1}>$, then a dataset that is randomly sampled from this buffer is used to train DQN.
			In this way, the DQL maximize the efficiency of learning experiences as one experience can be learned multiple times.			
		\begin{figure}[t]
			\centering
			\includegraphics[width=0.9\linewidth]{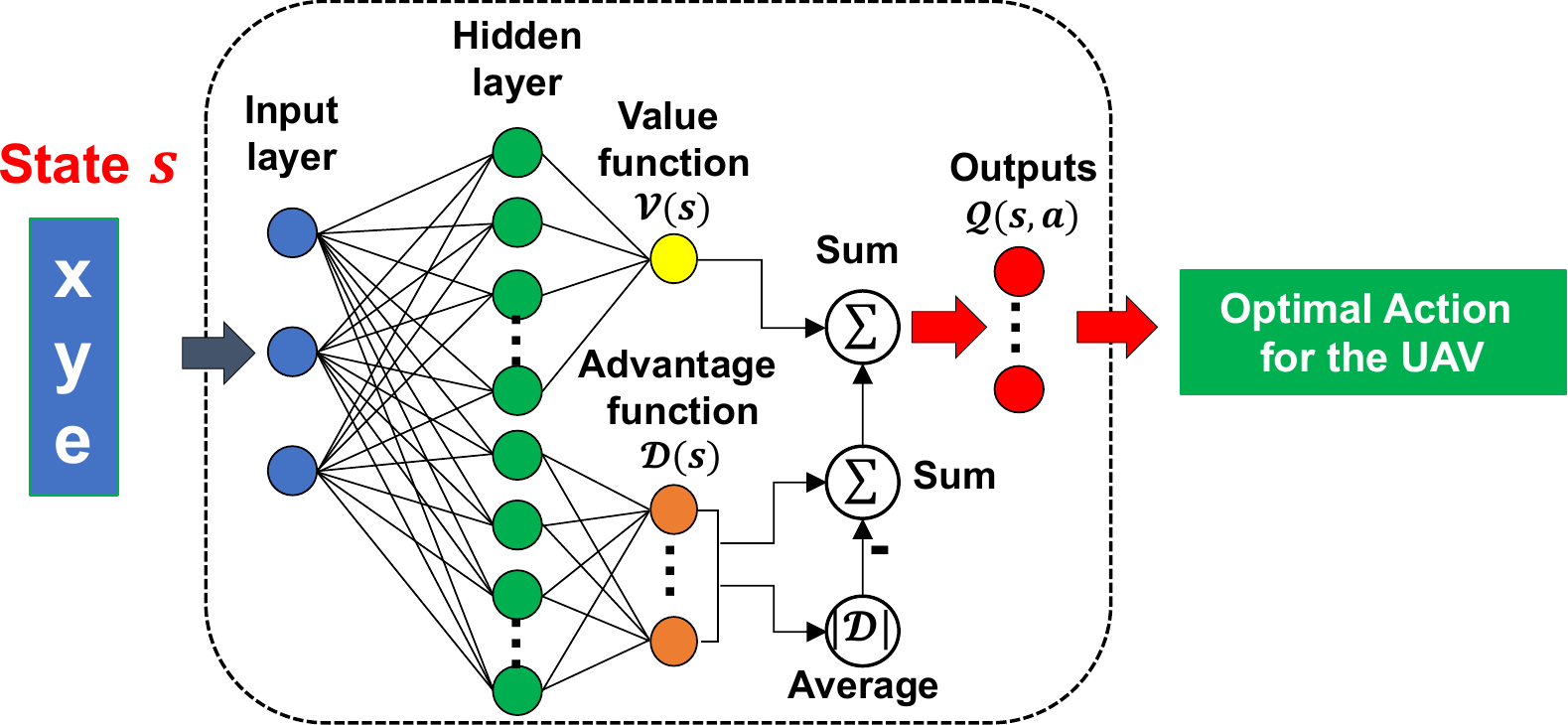}
			\caption{Deep dueling architecture~\cite{Wang2016Dueling}.}
			\label{fig.dueling_architecture}
		\end{figure}
						
		To improve the efficiency and stability of the learning process, the authors in~\cite{Wang2016Dueling} propose a dueling neural network architecture that consists of two parallel streams to simultaneously estimate the state-value function, i.e., $\mathcal{V}$, and the advantage function, denoted by $\mathcal{D}$, as illustrated in Fig.~\ref{fig.dueling_architecture}.	
			Recall that the state-value function $\mathcal{V}(s)$ specifies a the quality of a state $s$ (i.e., how good to be at a state), while the action-value function $\mathcal{Q}(s,a)$ determines the value of performing an action $a$ at a state $s$. 	
			The advantage function is defined by the relation between state-value and action-value functions under policy $\pi$ as follows~\cite{Wang2016Dueling}:
			\begin{equation}
				\label{eq:advantage_function}
				\mathcal{D}^\pi(s,a) = \mathcal{Q}^\pi(s,a) - \mathcal{V}^\pi(s).
			\end{equation}
			In this way, the advantage function $\mathcal{D}^\pi(s,a)$ indicates the importance of action $a$ in comparison with other actions at state $s$.
			Suppose that a state $s$ is inputted to the dueling neural network, the Q-function is then calculated by 
			\begin{equation}
				\label{eq:recontruct_Qfunction}
				\mathcal{Q}(s,a; \xi,\psi) = \mathcal{V}(s; \xi) + \mathcal{D}(s,a; \psi),		
			\end{equation}
			where $\xi$ and $\psi$ are weights and biases of the value and advantage streams, respectively. 
			It is worth noting that directly using \eqref{eq:recontruct_Qfunction} may lead to mediocre performance of the algorithm since this equation is unidentifiable, i.e., $\mathcal{V}$ and $\mathcal{D}$ cannot be uniquely determined by a given $\mathcal{Q}$.
			For example, if $\mathcal{V}$ increases the same amount that $\mathcal{D}$ decreases, $\mathcal{Q}$ is unchanged.
			In~\cite{Wang2016Dueling}, the authors address this issue by subtracting the maximum value of the advantage stream as follows:
			\begin{equation}
				\label{eq:recontruct_Qfunction_max}
				\mathcal{Q}(s,a; \xi,\psi) = \mathcal{V}(s; \xi) + \Big(\mathcal{D}(s,a; \psi) - \max_{a' \in \mathcal{A}} \mathcal{D}(s,a';\psi)\Big).		
			\end{equation}
			In this manner, $\mathcal{Q}(s,a^*; \xi,\psi)$ equals to  $\mathcal{V}(s; \xi)$ for the optimal action $a^*$, i.e., 
			\[a^* =  \text{argmax}_{a \in \mathcal{A}}~\mathcal{Q}(s,a; \xi,\psi) = \text{argmax}_{a \in \mathcal{A}}~\mathcal{D}(s,a; \psi).\]
			However, in \eqref{eq:recontruct_Qfunction_max}, the advantage stream changes as fast as the change in predicted optimal action's advantage, i.e., $\max_{a' \in \mathcal{A}} \mathcal{D}(s,a';\psi)$, leading to instability in estimating the Q-values. 
			Hence, we can replace the max operator by the mean to stabilize the estimation of Q-function as follows~\cite{Wang2016Dueling}:
			\begin{equation}
				\label{eq:recontruct_Qfunction_mean}
				\mathcal{Q}(s,a; \xi,\psi) = \mathcal{V}(s; \xi) + \Big(\mathcal{D}(s,a; \psi) - \frac{1}{|\mathcal{A}|} \sum_{a'}\mathcal{D}(s,a';\psi)\Big).		
			\end{equation}
	
		In addition, the Q-learning and deep Q-learning algorithms may overestimate Q-values, yielding a low performance in stochastic MDP \cite{Hasselt2016}. 
			Consequently, the overestimation of Q-values can negatively impact the learning process, even resulting in a sub-optimal policy if they are unevenly distributed over the state space~\cite{Thrun1993}. 
			To address the overestimation, double deep Q-learning (DDQL) is proposed in \cite{Hasselt2016}. 
			In DDQL, the max operator in \eqref{eq:updateQfunction} is decoupled into action selection and action evaluation, which are derived by two deep neural network-based estimators.
			To that end, we propose the D3QL algorithm that can take all the advantages of deep Q-learning, dueling architecture, and DDQL to increase the convergence rate, stabilize the learning process, and decrease the overestimation. 
			Thus, D3QL can quickly achieve the optimal operation, and thereby maximizing both data collection and energy efficiency for the UAV.
		\begin{figure}[t]
			\centering
			\includegraphics[width=0.9\linewidth]{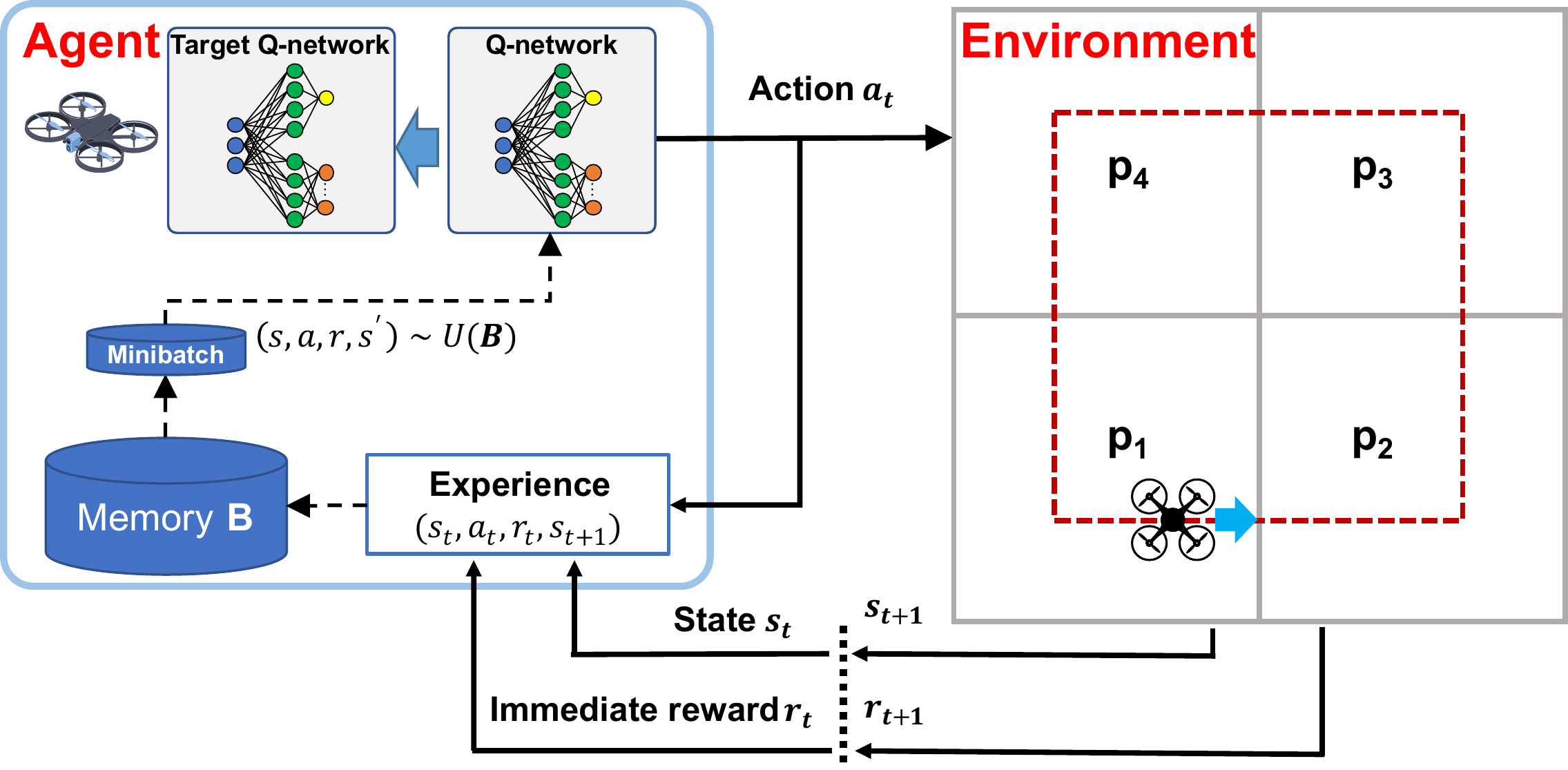}
			\caption{The proposed D3QL-based model.}
		\end{figure}	 
		\begin{algorithm}[t]
			\caption{The D3QL Algorithm}
			\label{alg:d3ql}
			\begin{algorithmic}[1]
				\STATE Establish  memory $\mathbf{B}$ and $\epsilon$. 
				\STATE Establish Q-network $\mathcal{Q}$ with random parameters $\phi$ and target Q-network $\hat{\mathcal{Q}}$ with parameters $\phi^-=\phi$.
				\FOR{\textit{step = 1 to T}}
				\STATE Choose action $a_t$ according to the $\epsilon$-greedy policy.
				\STATE Execute $a_t$, observe reward $r_t$ and next state $s_{t+1}$.
				\STATE Save experience $(s_t, a_t, r_t, s_{t+1})$ in  $\mathbf{B}$.
				\STATE Sample  mini-batch of experiences randomly from $\mathbf{B}$, i.e., $(s, a, r, s') \sim U( \mathbf{B}$).				
				\STATE Calculate Q-value $\mathcal{Q}(s_k, a_k; \phi)$ and target Q-value $G_t$ by~\eqref{eq:recontruct_Qfunction} and~\eqref{eq:targetDQN}, respectively.
				%				$Y_k = r_k + \zeta \hat{\mathcal{Q}}\big(s_{k+1},\underset{a}{\operatorname{argmax}}\mathcal{Q}(s_{k+1},a; \xi,\psi);\xi^-,\psi^-\big)$.}		   
				\STATE Take a gradient descent step with respect to the parameters of Q-network. %on $\big(Y_k-\mathcal{Q}(s_k, a_k; \phi)\big)^2$.
				\STATE Decrease the value of $\epsilon$.
				\STATE Set  $\hat{\mathcal{Q}}= \mathcal{Q}$ at every $I$ steps.
				\ENDFOR
			\end{algorithmic}
		\end{algorithm}
		\begin{figure*}[t]
			\centering
			$\begin{array}{cc}
				\includegraphics[width=0.4\linewidth]{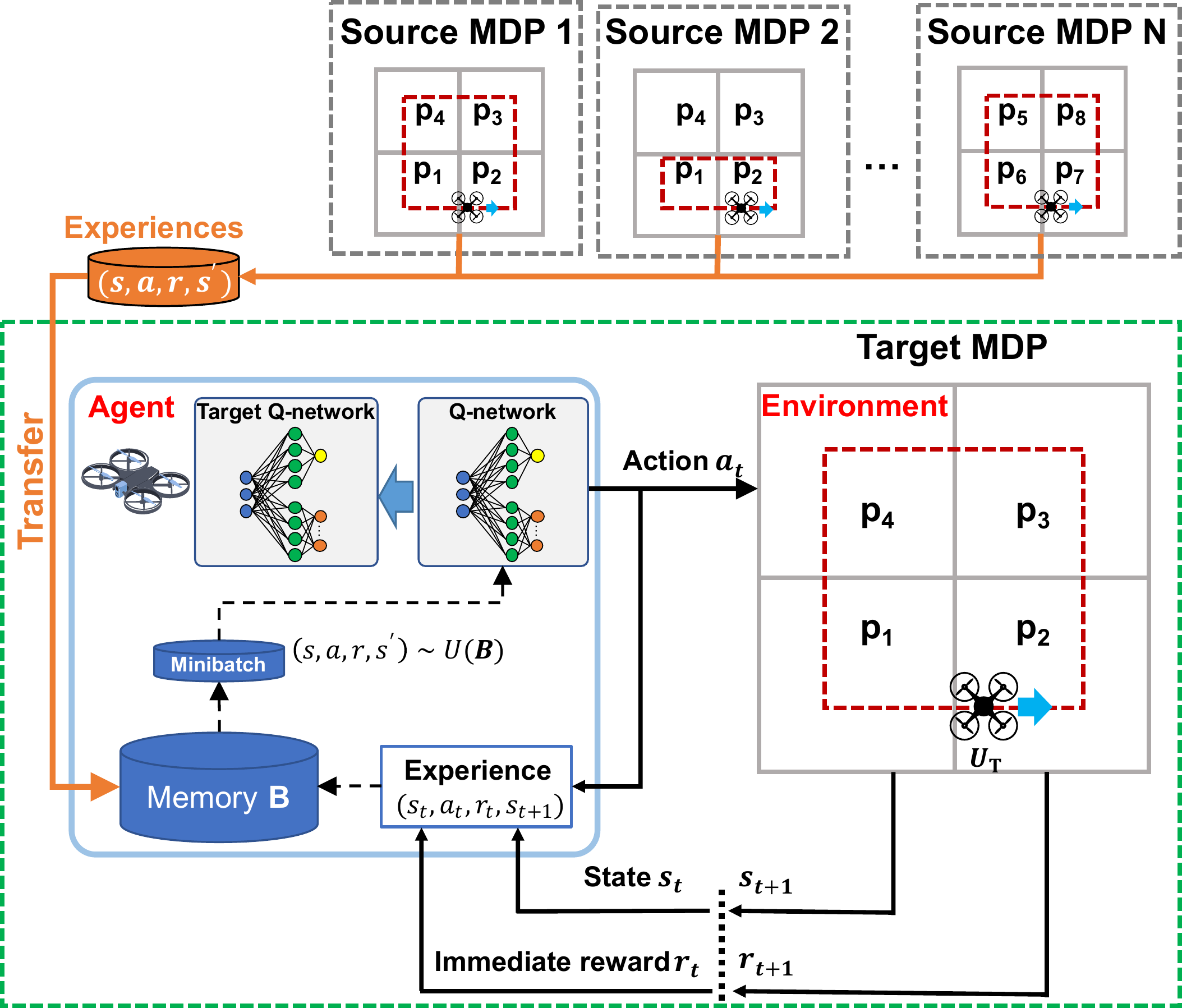} & \qquad
				\includegraphics[width=0.4\linewidth]{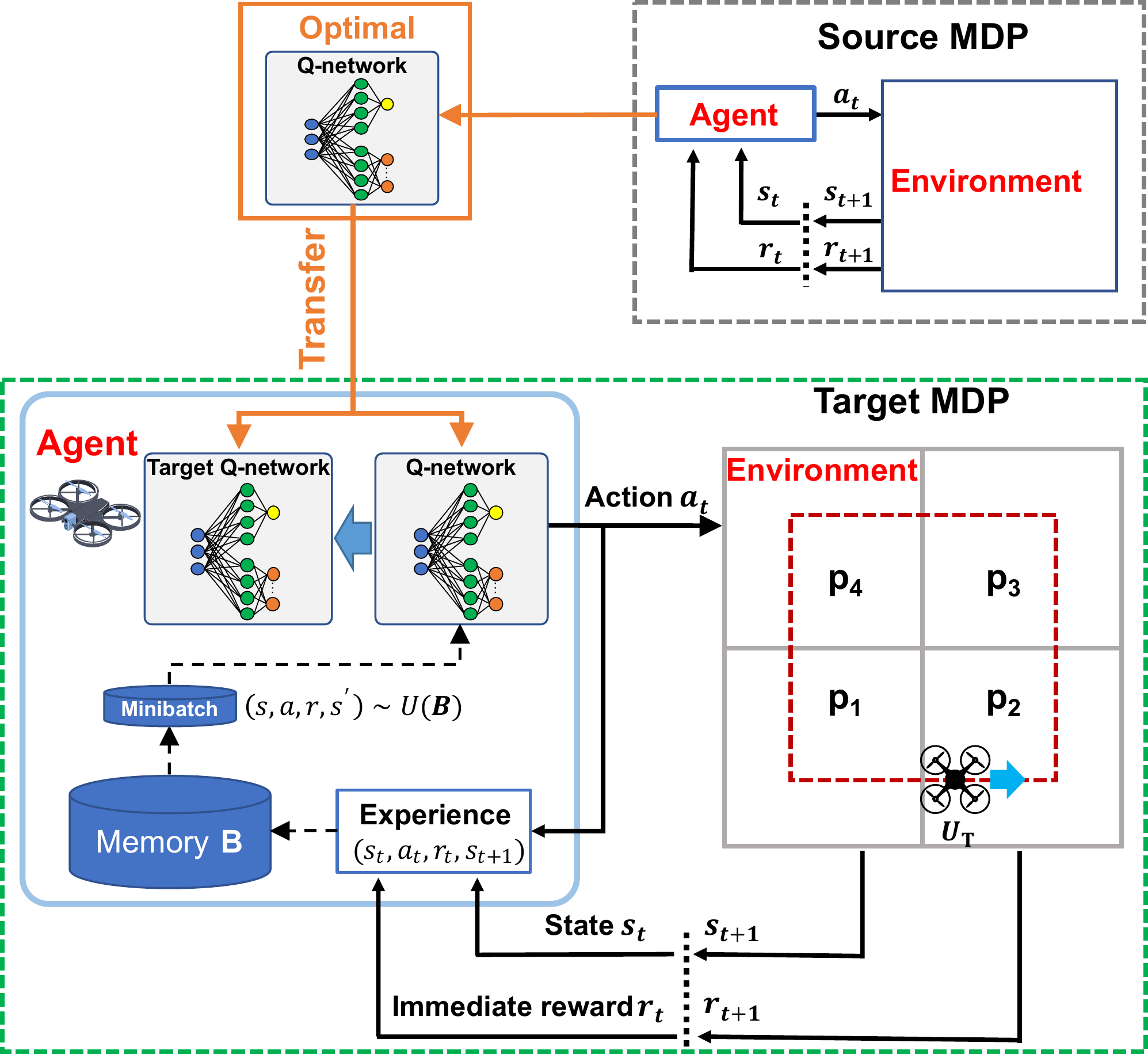} \\
				\text{(a) Experience transfer} & \qquad
				\text{(b) Policy transfer}
			\end{array}$
			\caption{The proposed D3QL-TL based models: (a) experience transfer and (b) policy transfer.}
			\label{fig.D3QL-TL_model}		
		\end{figure*}	
		The proposed D3QL algorithm is thoroughly described in Algorithm~\ref{alg:d3ql}. Specifically, the learning process consists of some major steps. 
			At the beginning of time step $t$, the UAV is at state $s_t$ and performs action $a_t$ according to $\epsilon$-greedy policy.   			
			It then observes the next state $s_{t+1}$ and receives reward $r_t$ at the end of a time slot. 
			We can use the experience, i.e., a tuple $<s_t,a_t,r_t,s_{t+1}>$, to train the neural network at step $t$.
			However, the high correlation in consecutive experiences generated by MDP can severely slow the learning process~\cite{halkjaer_the_1996}.
			To that end, D3QL stores the experiences in a memory $\mathbf{B}$, then samples a mini-batch uniformly from this memory, resulting in a decrease of the variance of the update, and thereby improving the convergence rate of the algorithm. 
			In addition, the usage of the memory pool increases the data efficiency of the network's training since one experience can be used to update the neural network multiple times.
			In the proposed D3QL, the input layer of the dueling network has three inputs corresponding to the UAV's state dimensions, i.e., the x-coordinate, y-coordinate, and current energy level.		
		
		The Q-learning's overestimation is handled in D3QL by employing two dueling neural network-based estimators, which are Q-network, i.e., $\mathcal{Q}(s,a; \xi,\psi)$, and target Q-network, i.e., $\hat{\mathcal{Q}}(s,a; \xi^-,\psi^-)$. 
			Specifically, the Q-network is used for action selection, while the target Q-network is used for action evaluation. Thus, the target Q-value at step $t$ is now defined by:
			\begin{equation}
				\label{eq:targetDQN}
				G_t = r_t + \zeta \hat{\mathcal{Q}}\big(s_{t+1}, \underset{a}{\operatorname{argmax}}\mathcal{Q}(s_{t+1},a; \xi_t, \psi_t);\xi^-_t, \psi^-_t\big).
			\end{equation}
%		Recall that the dueling neural network parameters consist of value stream $\xi$ and advantage streams $\psi$ parameters. 
		For convenience, let $\phi_t$ and $\phi^-_t$ denote the parameters of Q-network and those of target Q-network at time $t$, respectively. 
		Since the goal of Q-network training is to minimize the TD, we can define the loss function at time $t$ as follows:
			\begin{equation}
				\begin{aligned}
					\label{eq:lossfunction}
					\mathcal{L}_t(\phi_t) = \mathbb{E}_{(s,a,r,s')}\bigg[ \bigg( G_t
					-\mathcal{Q}(s,a;\phi_t)\bigg)^2\bigg],
				\end{aligned}
			\end{equation}
			where $(s,a,r,s')$ represents a data point in memory $\mathbf{B}$ used to train the Q-network.
		
		Gradient Descent (GD) algorithm is widely used to minimize the loss function for deep learning algorithms because of its simplicity in implementation and ability to find the global minimal~\cite{du_gradient_2019}. 
			In particular, GD calculates the cost function at time $t$ as follows:
			 \begin{equation}
			 	\begin{aligned}
			 		\label{eq:costfunction}
			 		J_t(\phi_t) = \frac{1}{|\mathbf{B}|}\sum_{(s,a,r,s') \in \mathbf{B}}\mathcal{L}_t(\phi_t).
			 	\end{aligned}
			 \end{equation}
			Then, the parameters of Q-network are updated by
			\begin{equation}
				\label{eq:GDupdate}
				\phi_{t+1} = \phi_t -\alpha_t \nabla_{\phi_t} J_t(\phi_t),
			\end{equation}
			where $\alpha_t$ is a step size at time $t$ determining how much the parameters are updated, and $\nabla_{\phi_t}(.)$ is the gradient of the cost function with respect to the parameters of Q-network $\phi$.
			For each update, GD requires computing the loss and gradient for all data points in the memory $\mathbf{B}$, leading to a very slow process if the data size is large, especially for the considered problem. 
			To that end, we adopt stochastic gradient descent (SGD) to speed up the parameter update process while guaranteeing the learning convergence~\cite{Robbins1951SGD}.
			Specifically, SGD only needs to calculate the gradient and cost of a mini-batch sampled uniformly at random from $\mathbf{B}$ for each time step.
			Hence, the computational complexity of this algorithm is significantly decreased.   
			It is worth noting that the target Q-values $G_t$ in the loss function~\eqref{eq:lossfunction} appear similar to the labels used for supervised learning. However, instead of being fixed before the training, $G_t$ changes as fast as the changes in the Q-target network's parameters $\phi^-_t$. 
			Therefore, the target Q-network's parameters are only updated by cloning from the Q-network's parameters at every $I$ steps to stabilize the training process.
								
	\subsection{Optimal Operation Policy with Transfer Learning}
	Although D3QL can effectively address the shortcomings of Q-learning, it still poses some drawbacks inherited from conventional DRL when addressing scenarios with high sample complexity, as the considered problem in this work where the surrounding environment of the UAV is unknown in advance.
		First, it often takes a lot of time to train DNN, e.g., DQN's training time is up to $38$ days for each Atari game~\cite{Mnih2015Human}.
		If the environment dynamics or the trajectory of the UAV changes, the DNN may need to be retrained from scratch, yielding a high computing complexity.
		Second, since the UAV should only return for replacing its battery when it is close to the station or its energy level is low, it needs sufficient experiences in this region, especially when flying over the station.
		However, as the UAV flies over its fixed trajectory, experiences obtained from this region are often very small compared with all obtained over the entire considered area. Therefore, the UAV may not have adequate information to learn an optimal policy.
		To address these challenges, we develop a novel D3QL framework leveraging transfer learning techniques, namely Deep Dueling Double Q-learning with Transfer Learning (D3QL-TL).
		
	\subsubsection{Transfer Learning in Reinforcement Learning}	
		Transfer learning is a method of leveraging knowledge obtained when performing a source task in a source domain to enhance the learning process of target tasks in target domains~\cite{taylor_transfer_2009, pan_survey_2009, zhu_trasfer_2020, nguyen_transfer_2021}. 
		Typically, a domain contains labeled or unlabeled data given before the considered training process starts. 
		However, data in RL is obtained via interactions between the agent (i.e., the UAV) and its surrounding environment. 
		As a result, both the domain and task can be represented by an MDP. 	
		Thus, we can define transfer learning in RL as in Definition~\ref{def:transfer_learning}~\cite{zhu_trasfer_2020}.
		\begin{definition}
			\label{def:transfer_learning}
			\begin{spacing}{1}
			 Transfer learning in RL: Suppose the source and target MDPs are defined. Transfer learning (TL) in RL intends to leverage the knowledge $\mathcal{K}_\mathrm{S}$ obtained from the source MDP, i.e., the policy, the environment dynamics, and the data,  as a supplement to the target MDP's information $\mathcal{K}_\mathrm{T}$ to efficiently learn the target optimal policy $\pi^*_\mathrm{T}$ as follows:
			\begin{equation}
				\pi^*_\mathrm{T} = \underset{\pi_\mathrm{T}}{\operatorname{argmax}} \mathbb{E}_{s \sim \mathcal{S}_\mathrm{T}, a\sim \pi_\mathrm{T}} \left[ \mathcal{Q}^{\pi_\mathrm{T}}(s,a)\right],
			\end{equation}	
		where $\pi_\mathrm{T}$ is a target MDP's policy approximated by an estimator, e.g., a table or a deep neural network, that is trained on both $\mathcal{K}_\mathrm{S}$ and $\mathcal{K}_\mathrm{T}$.			
					\end{spacing}	
	\end{definition}
		To measure the effectiveness of TL, we can use three metrics, including jump-start, asymptotic performance, and time-to-threshold~\cite{taylor_transfer_2009}. 
		In particular, jump-start measures how much the UAV's performance at the beginning of the learning process can be improved by applying TL, while the asymptotic performance measures this improvement at the end of the learning process.
		The third metric, i.e., time-to-threshold, measures how fast TL can help the UAV to achieve a predefined performance level compared with the scenario without TL.
		It is worth highlighting that TL cannot guarantee improvement in the learning curve. 
		It may even negatively impact the learning in the target MDP if the transfer knowledge is not carefully chosen. 
		Thus, in the following sections, we propose a transfer learning framework that can reduce the learning time and learning quality for D3QL.
	
	\subsubsection{Deep Dueling Double Q-learning with Transfer Learning}
	The details of D3QL-TL are presented in Algorithm~\ref{alg:d3ql-tl}.
		In particular, as illustrated in Fig.~\ref{fig.D3QL-TL_model}, we consider a UAV $U_\mathrm{S}$ working in an IoT data collection environment formulated by the MDP framework $\mathcal{M}_\mathrm{S}$. 
		Then, the knowledge of $U_\mathrm{S}$ can be leveraged to help a new UAV $U_\mathrm{T}$ effectively learn the optimal policy for working in another environment formulated by the MDP framework $\mathcal{M}_\mathrm{T}$.
		In practice, $\mathcal{M}_\mathrm{T}$ can be the same or different with $\mathcal{M}_\mathrm{S}$.
		Moreover, the transferring knowledge can be in the form of the policy and/or experiences of the source UAV, i.e.,  $U_\mathrm{S}$.
		\begin{algorithm}[t]
			\caption{The D3QL-TL}
			\label{alg:d3ql-tl}
			\begin{algorithmic}[1]
				\STATE Establish  memory $\mathbf{B}$ and $\epsilon$. 
				\STATE Establish Q-network $\mathcal{Q}$ with random parameters $\phi$, and target Q-network $\hat{\mathcal{Q}}$ with parameters $\phi^-=\phi$.
				\IF{\text{Experience transfer}}
				\STATE Copy the experience set of source MDP $\mathcal{M}_{\mathrm{S}}$ to $\mathbf{B}$.
				\ELSIF{\text{Policy transfer}}	
				\STATE Re-initialize $\mathcal{Q}$ and $\hat{\mathcal{Q}}$ with the parameter of source Q-network's parameters $\phi_{\mathrm{S}}$.
				\ELSIF{\text{Hybrid transfer}}	
				\STATE Re-initialize $\mathcal{Q}$ and $\hat{\mathcal{Q}}$ with the parameter of source Q-network's parameters $\phi_{\mathrm{S}}$.
				\STATE Copy the experience set of source MDP $\mathcal{M}_{\mathrm{S}}$ to $\mathbf{B}$.
				\ENDIF
				\FOR{\textit{step = 1 to T}}
				\STATE Choose action $a_t$ according to the $\epsilon$-greedy policy.
				\STATE Execute $a_t$, observe reward $r_t$ and next state $s_{t+1}$.
				\STATE Save experience $(s_t, a_t, r_t, s_{t+1})$ in $\mathbf{B}$.
				\STATE Sample  mini-batch of experiences randomly from $\mathbf{B}$, i.e., $(s, a, r, s') \sim U( \mathbf{B}$).				
				\STATE Calculate Q-value $\mathcal{Q}(s_k, a_k; \phi)$ and target Q-value $G_t$ by~\eqref{eq:recontruct_Qfunction} and~\eqref{eq:targetDQN}, respectively.
				%				$Y_k = r_k + \zeta \hat{\mathcal{Q}}\big(s_{k+1},\underset{a}{\operatorname{argmax}}\mathcal{Q}(s_{k+1},a; \xi,\psi);\xi^-,\psi^-\big)$.}		   
				\STATE Take a gradient descent step with respect to the parameters of Q-network. %on $\big(Y_k-\mathcal{Q}(s_k, a_k; \phi)\big)^2$.
				\STATE Decrease the value of $\epsilon$.
				\STATE Set  $\hat{\mathcal{Q}}= \mathcal{Q}$ at every $I$ steps.
				\ENDFOR
			\end{algorithmic}
		\end{algorithm}
		
		To that end, D3QL-TL defines three types of knowledge transferring as follows:
		\begin{itemize}
			\item \textit{Experience Transfer (ET)}: This approach aims to leverage a set of experiences $E_\mathrm{S}$, in which each element is a experience tuple $<s,a,r,s'>$, obtained in the source MDP, i.e., $\mathcal{M}_\mathrm{S}$, to improve the learning process of the target UAV, i.e., $U_\mathrm{T}$, working in the target MDP, i.e., $\mathcal{M}_\mathrm{T}$. 
				Specifically, $E_\mathrm{S}$ is first copied to the memory buffer of the target UAV. 
				Then, these transferred experiences and target UAV's new experiences are used to train the Q-network. 
				In this manner, the target UAV can quickly get adequate information, and thereby significantly  improving the learning speed. 
				In addition, the quality of the experiences also affects the learning process. 
				For example, an experience does not have much value if it is easy to be obtained by the target UAV. 
				In contrast, an experience is considered to be valuable if it is hard to obtain and highly impacts the system performance.
				For example, experiences obtained when the UAV is near the station may have high values because they not only contain information about environment dynamics (i.e., packet arrival probabilities) but also may reveal value information about the right time to take the battery replacement action.				
			\item \textit{Policy Transfer (PT)}: This approach directly transfers the policy of a source UAV to a target UAV. 
				In particular, $U_\mathrm{T}$ starts the learning process with the policy of $U_\mathrm{S}$, which is represented by the Q-network of $U_\mathrm{S}$, called the source Q-network. 
				Hence, the $U_\mathrm{T}$'s Q-network is initialized by the source Q-network parameters $\phi_{\mathrm{S}}$. 
				Then, the $U_\mathrm{T}$'s Q-network is trained with the new experiences of $U_\mathrm{T}$ obtained in the target MDP $\mathcal{M}_\mathrm{T}$.
				Thus, starting with the source policy can help $U_\mathrm{T}$ to avoid random decisions caused by the randomness of action selection at the beginning of the learning process, e.g., inappropriately choosing the battery replacement action.
			\item \textit{Hybrid}: This approach aims to leverage the benefits of both experience and policy transfer types.
				Particularly, the hybrid scheme can improve not only the jump-start but also the asymptotic performance. 
		\end{itemize}
	Note that the efficiency of each transferring technique depends on the relationship between the source and the target MDPs. For example, if the source and target MDPs are very similar, policy transfer may yield a better result in terms of convergence rate than that of the experience transfer technique. In contrast, when the source and target MDP are not similar, e.g., differences in environment dynamics, the experience transfer should be a better choice. We will explore these techniques through intensive simulation results in the next section.
	
%
%%--------------------------------------------------------------------------------------------------------------------------
%%--------------------------------------------------------------------------------------------------------------------------	
\section{Performance Evaluation}
\label{sec:PE}
	\subsection{Parameter Setting}
	\label{subsec:PS}
	\begin{figure}[t]
		\centering
		$\begin{array}{ccc}
			\includegraphics[width=0.3\linewidth]{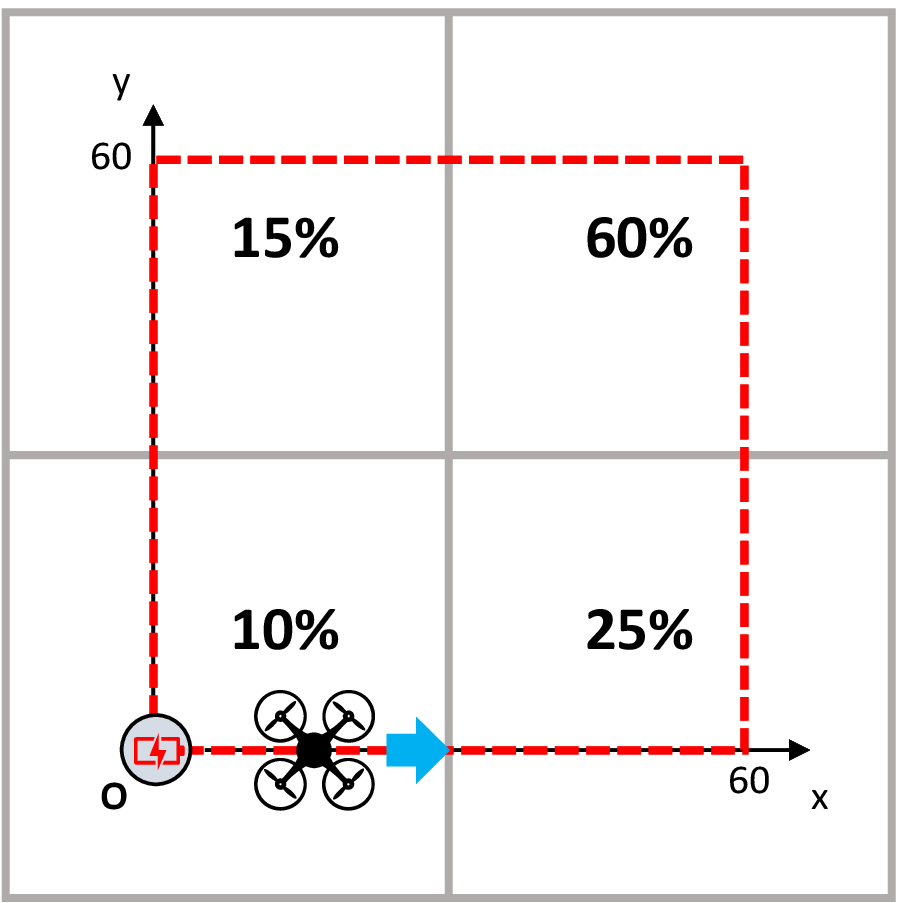} &
			\includegraphics[width=0.3\linewidth]{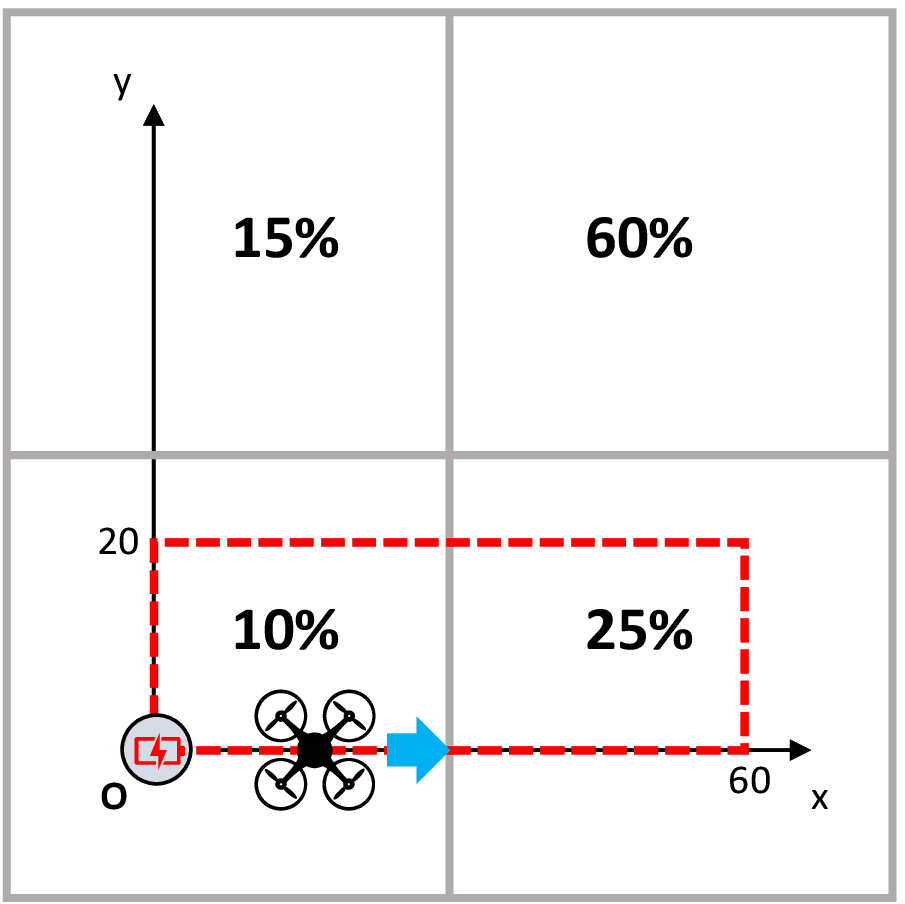} &
			\includegraphics[width=0.3\linewidth]{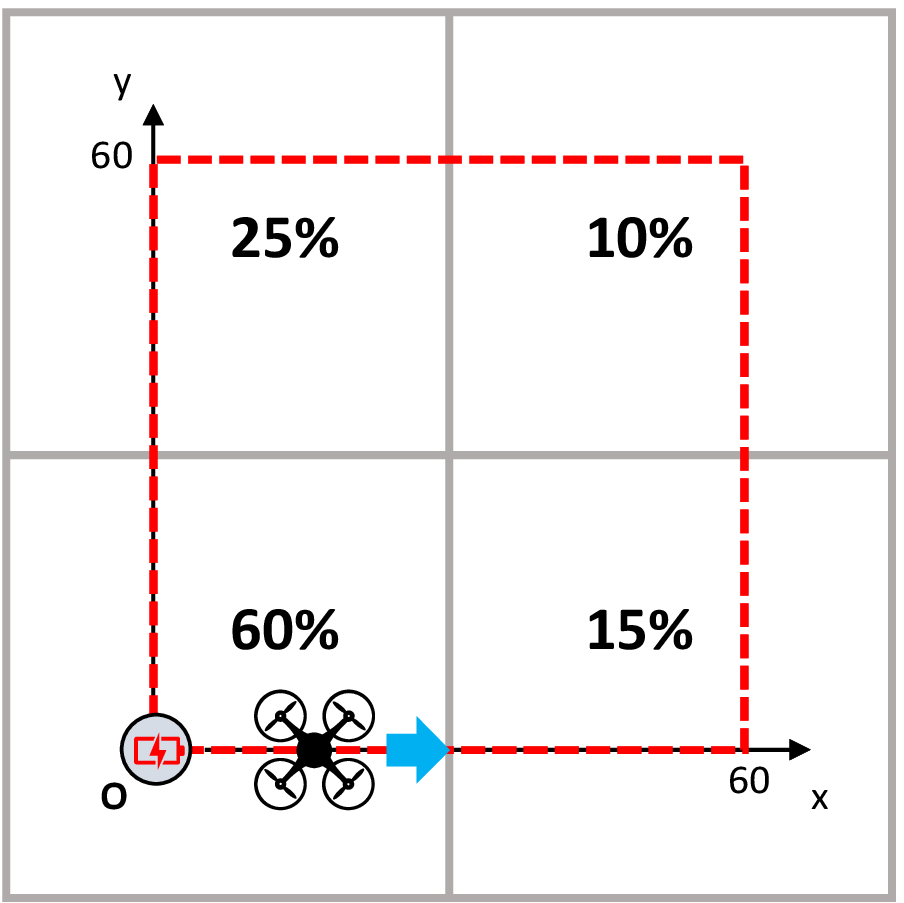} \\
			\text{(a) $\mathcal{M}_\mathrm{S}$}&
			\text{(b) $\mathcal{M}_\mathrm{T}^1$} &
			\text{(c) $\mathcal{M}_\mathrm{T}^2$}				
		\end{array}$
		\caption{(a) Source MDP, (b) the first target MDP, and (c) the second target MDP.}
		\label{fig.simulation_MDPs}
	\end{figure}	
	
	\begin{table*}[t]
		\centering
		\begin{tabular}{|c|c|c|c|c|c|c|c|c|c|c|c|c|c|}
			\hline
			\textbf{Parameters}	& $\Omega$ & $w_1$ & $w_2$ & $c_1$ & $c_2$ & $c_3$ & $c_4$ & $c_5$ & $E$ & $\mathbf{p}$ & $t_{b}$ & $v_r$ \\
			\hline
			\textbf{Value} 		& $1$	  & $1$			& $0.3226$   & $5$  &  $0.5$ & $0.5$ & $0.022$ & $0.2$ &  $300$   & $[0.1,0.25,0.6,0.15]$	 & $10$  & $1$ \\
			\hline
		\end{tabular}	
		\caption{Simulation parameters.}
		\label{table:simulation_parameters}		
	\end{table*}

	\begin{figure}[t]
		\centering
		$\begin{array}{c}
			\includegraphics[width=0.7\linewidth]{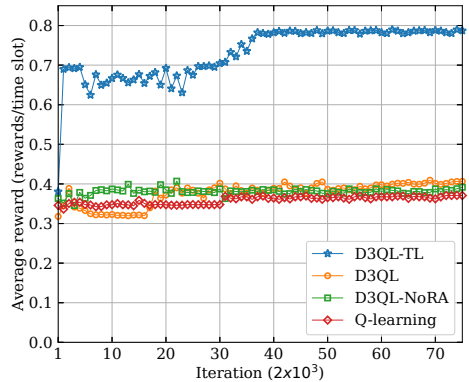} \\
			\text{(a) Convergence rate of proposed algorithms} \\
			\includegraphics[width=0.7\linewidth]{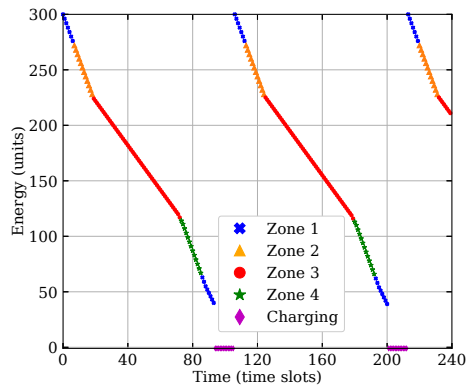} \\
			\text{(b) Policy obtained by D3QL-TL}
		\end{array}$
		\caption{Convergence rate and policy.}
		\label{fig:convergene_policy}
	\end{figure}	
	We first evaluate our proposed approaches in an IoT system, in which a UAV flies over a predefined trajectory in a considered area to collect IoT data. 
		This area is divided into four zones such that the UAV's travel distance in each zone equals $60m$, as illustrated in Fig.~\ref{fig.simulation_MDPs}(a). 
		The station is located at the origin, i.e., $(0,0)$.
		The probabilities of packet arrival in these zones are given by a vector $\mathbf{p} = [p_1,p_2,p_3,p_4]$, e.g., $p_1$ is the packet arrival probability when the UAV is flying over zone $1$ and so on. 	
		Since the UAV collects data while flying, it often flies at a low speed (e.g., $5~m/s$) to maintain the reliability of the data collection process~\cite{Pan2018Sensor}. 
		Therefore, we consider that the UAV has three speeds: $1,3,\text{ and }5~(m/s)$.
		The UAV consumes $2,3,\text{ and }4$ (energy units/time slot) when flying with these speeds, respectively.
		Note that these energy consumption values are only to demonstrate how much energy the UAV uses for operation.
		In practice, UAVs manufactured by different brands may have different specifications.  
		The parameters of the reward function are provided in Table.~\ref{table:simulation_parameters}.		
		
	The settings for our proposed algorithms are set as follows. 
		For the $\epsilon$-greedy policy, $\epsilon$ is first set at $1$, then gradually decreased to $0.01$. 
		For all the proposed algorithms, the discount factor is set at $0.9$. In Q-learning, the learning rate $\beta$ is $0.1$. 
		The architectures of Q-network and target Q-network are illustrated in Fig.~\ref{fig.dueling_architecture}. 
		We use typical hyperparameters for training DNN, e.g., the learning rate and the frequency update of $\hat{\mathcal{Q}}$ are set to $10^{-4}$ and $10^4$, respectively, as those in~\cite{Mnih2015Human, Goodfellow2016Deep}. 

	For the experience transfer approach in D3QL-TL, experiences are selected to be transferred based on their valuable information. 
		Recall that if the UAV performs the battery replacement action when it is far from the station, it always receives a very small reward compared with those of other actions, as in \eqref{eq:total_reward_function}. 
		Therefore, the UAV's experiences in this area can imply that it should not take the battery replacement action.
		In contrast, experiences obtained when the UAV is near the station can contain both information, i.e., when it not worth to return for charging (e.g., current energy level is high) and when it is worth to take the battery replacement action (e.g., current energy level is low).
		Thus, we choose experiences obtained when the UAV flies near the station to be transferred.	    
	
	We study a scheme where the UAV does not have complete information about the surrounding environment in advance, e.g., the data arrival probabilities and the energy replenishment process. 
		Hence, we compare our approach with three other deterministic policies, i.e., the UAV always flies at (1) lowest speed ($1m/s$), (2) middle speed ($3m/s$), and (3) highest speed ($5m/s$). 
		Moreover, to investigate impacts of the battery replacement action on the system performance, we consider an approach, namely D3QL-NoRA, where the D3QL will still be implemented on the UAV, but the UAV will not select the battery replacement action.
	
	\subsection{Simulation Results}
	In the simulation, we first evaluate the performance of our proposed learning algorithm, i.e., D3QL-TL, by examining  the convergence rate and the obtained policy.				
		Then, we evaluate the system performance when varying some important parameters (e.g., battery replacement time, UAV's energy capacity, return speed, and packet arrival probabilities) to assess their influences on the system performance.
		For the D3QL-TL, the experience transfer type is chosen because it can leverage the experiences obtained during the learning phases of other algorithms, i.e., D3QL and Q-learning.
		Finally, to gain more insights into the effectiveness of three transferring types in D3QL-TL, we compare their performance in different scenarios, i.e., changing the UAV trajectory and the probabilities of receiving a packet.

		\subsubsection{Convergence and Policy}
		\label{subsubsec:convergence}
			\begin{figure*}[t]
				\centering
				\begin{subfigure}[b]{0.3\textwidth}
					\centering
					\includegraphics[scale=0.2]{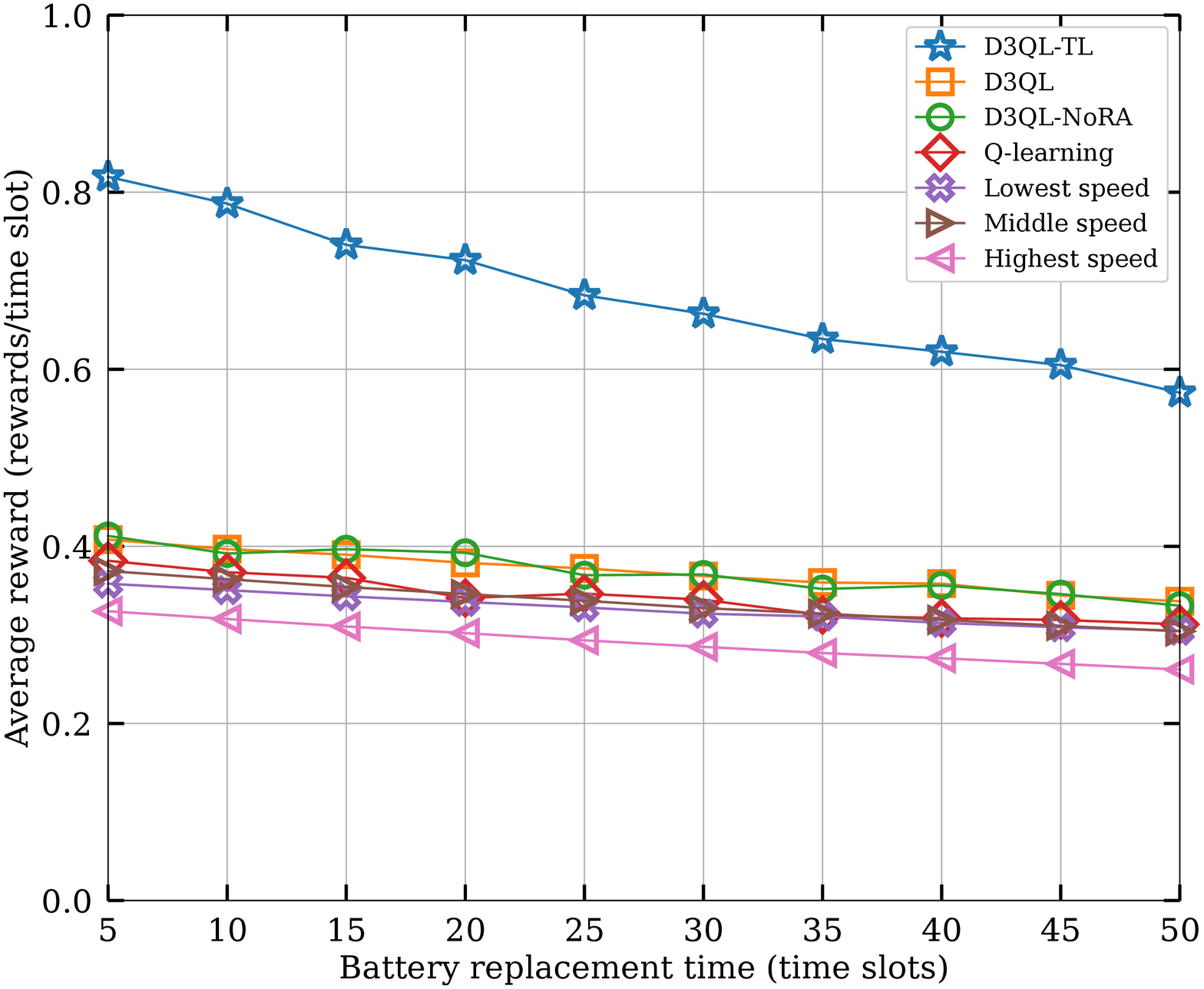}
					\caption{Average reward}
				\end{subfigure}
				~ 
				\begin{subfigure}[b]{0.3\textwidth}
					\centering
					\includegraphics[scale=0.2]{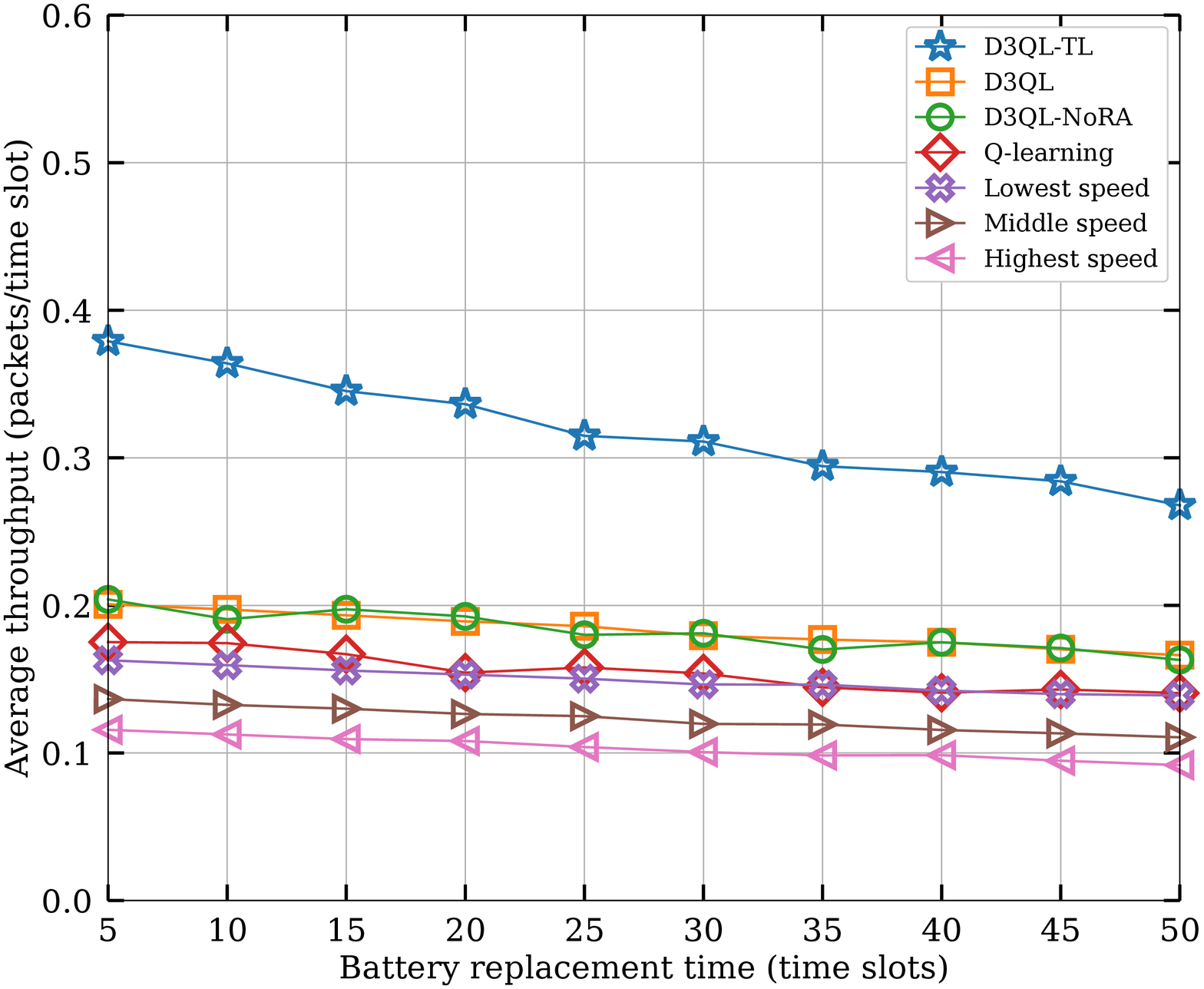}
					\caption{Average throughput}
				\end{subfigure}
				~
				\begin{subfigure}[b]{0.3\textwidth}
					\centering
					\includegraphics[scale=0.2]{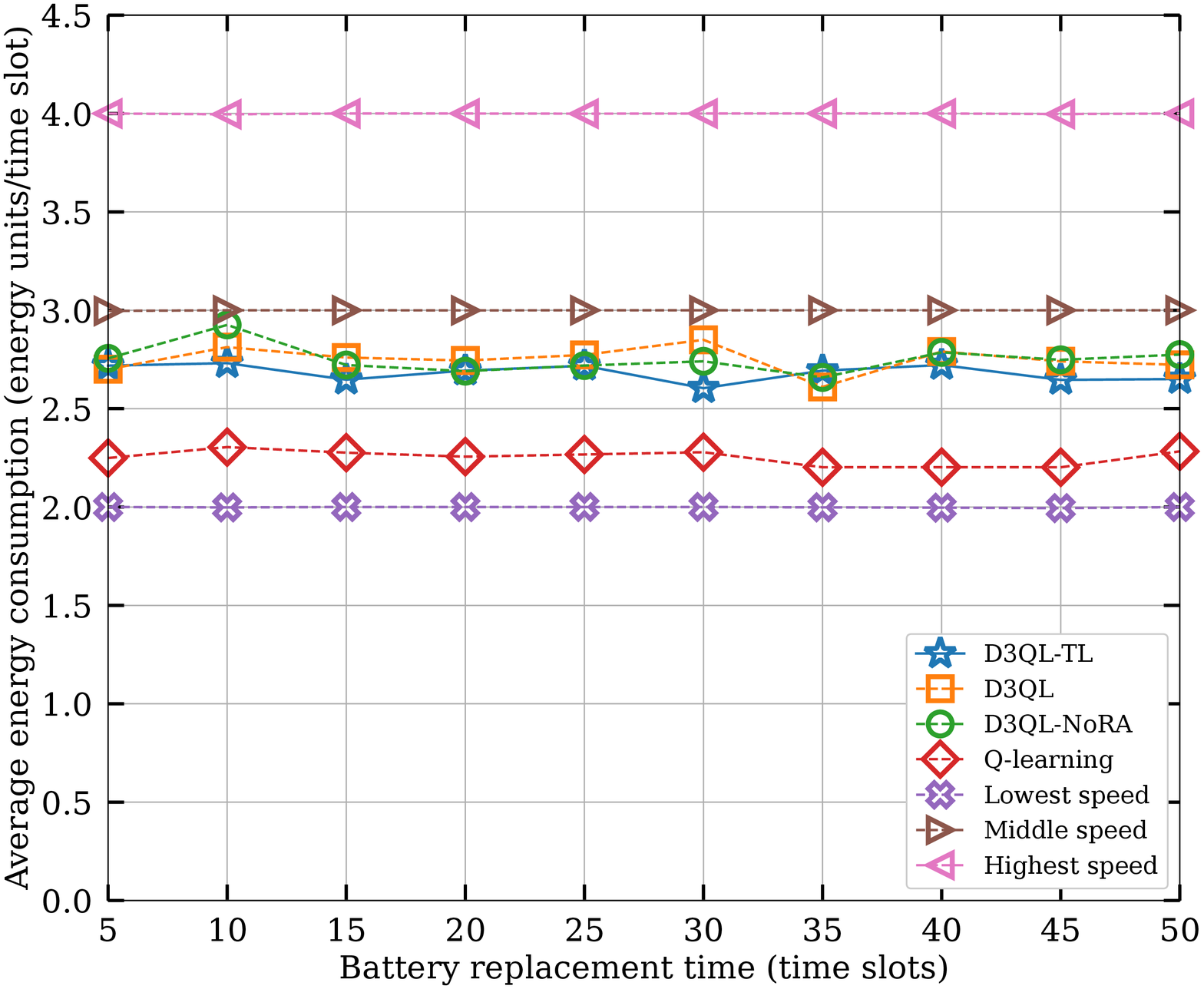}
					\caption{Average energy consumption}
				\end{subfigure}					
				\caption{Vary battery replacement time.} 
				\label{fig:vary_battery_replacement}
			\end{figure*}						
			\begin{figure*}
				\centering
				\begin{subfigure}[b]{0.3\textwidth}
					\centering
					\includegraphics[scale=0.2]{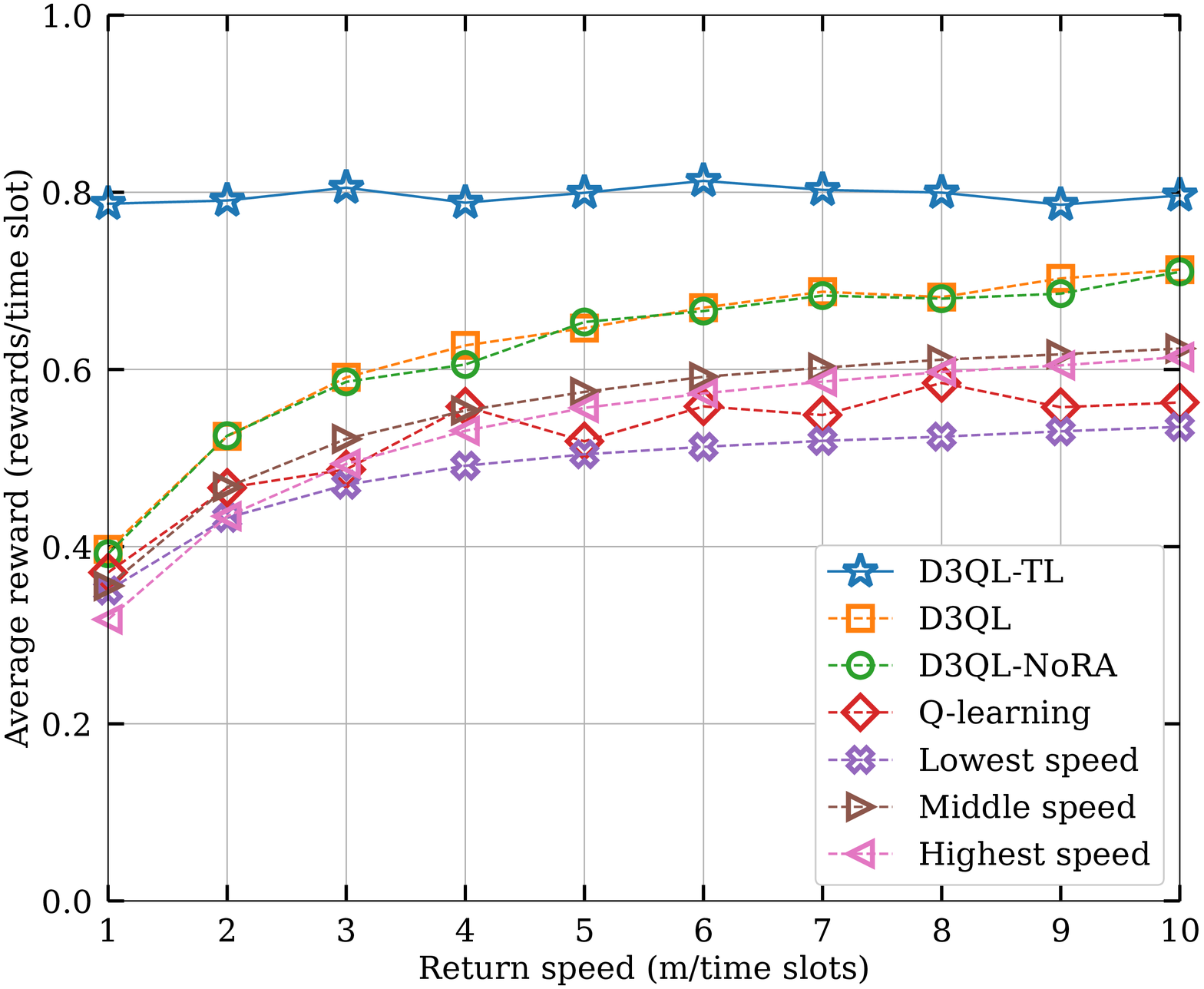}
					\caption{Average reward}
				\end{subfigure}%
				~ 
				\begin{subfigure}[b]{0.3\textwidth}
					\centering
					\includegraphics[scale=0.2]{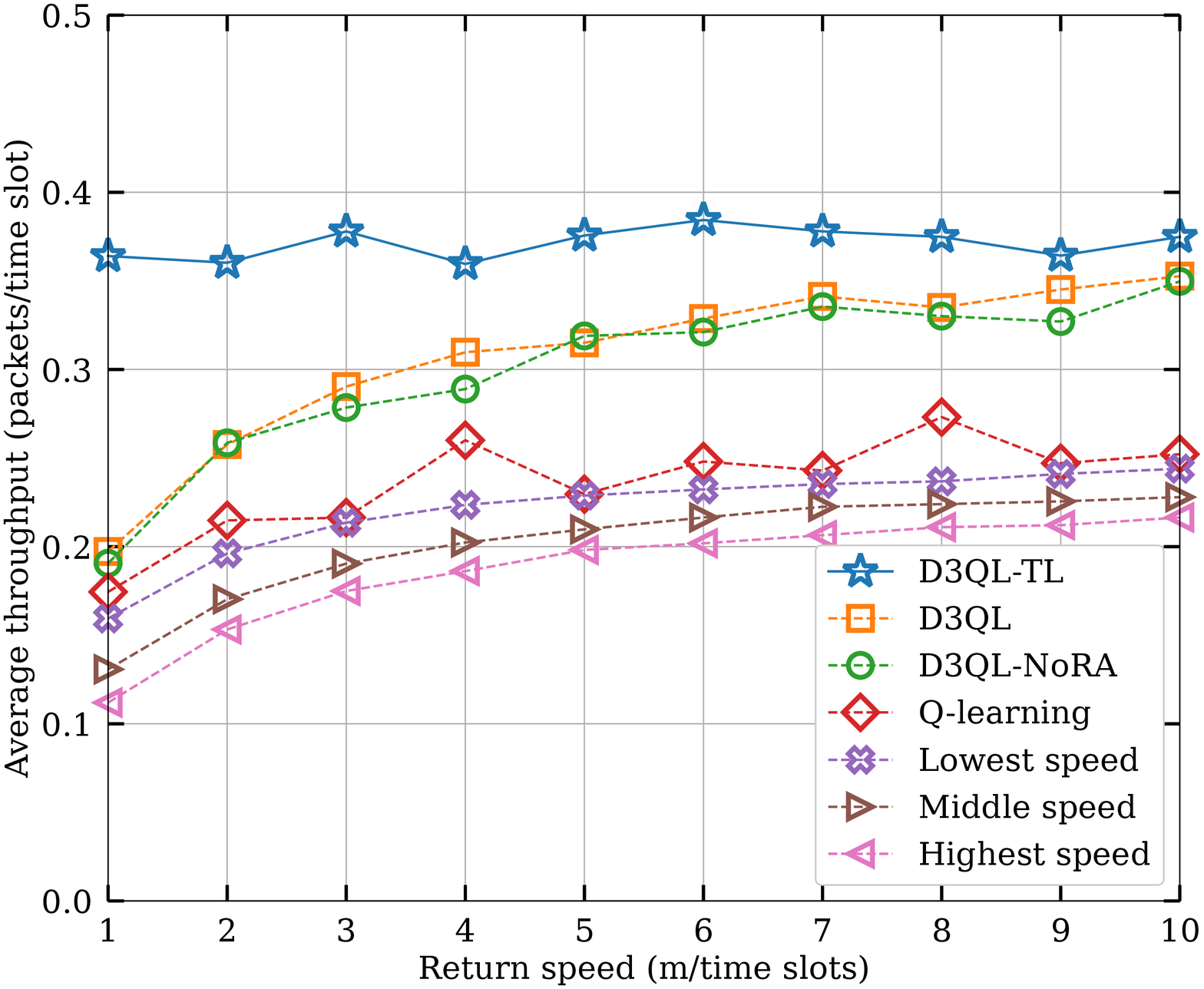}
					\caption{Average throughput}
				\end{subfigure}
				~
				\begin{subfigure}[b]{0.3\textwidth}
					\centering
					\includegraphics[scale=0.2]{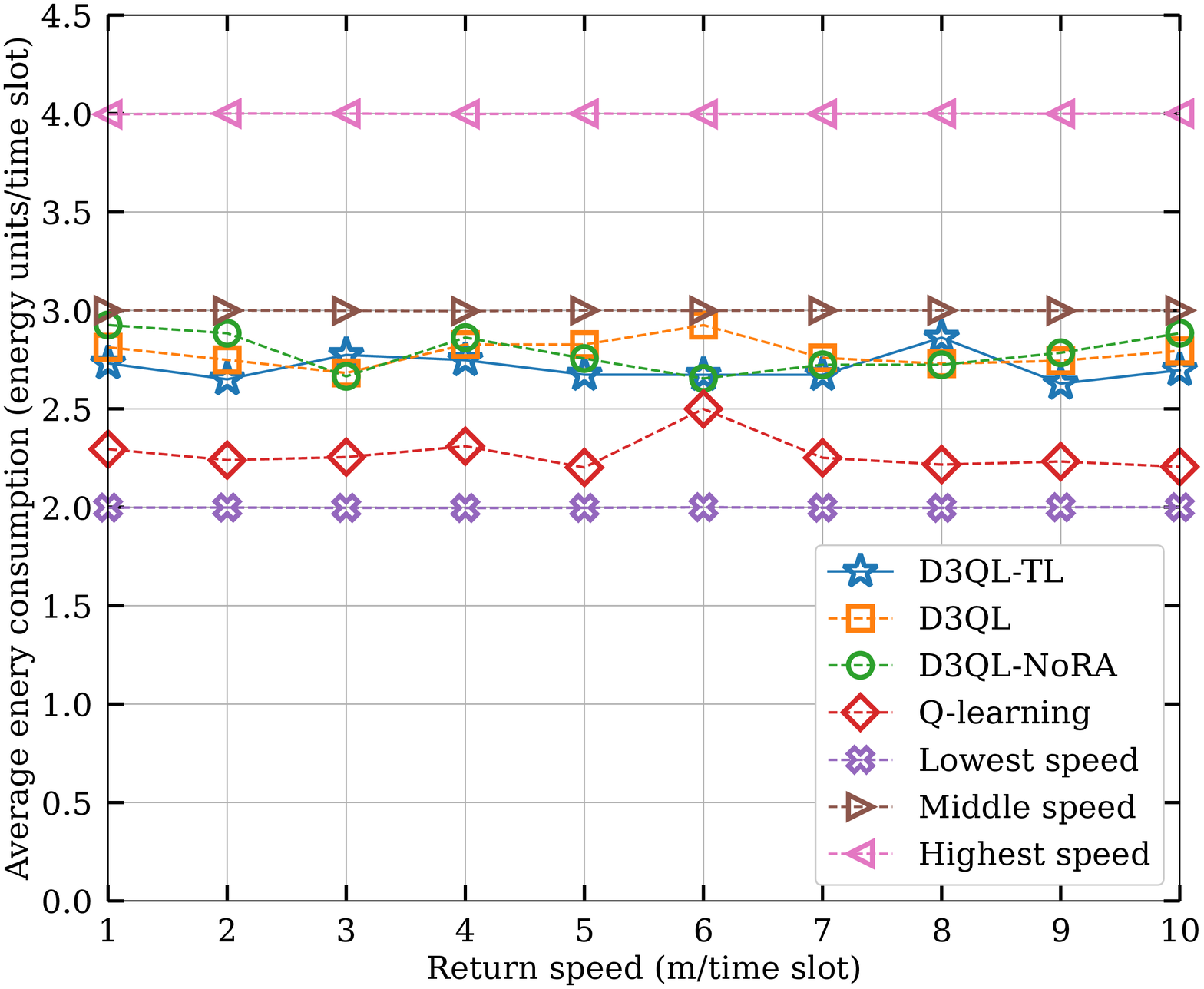}
					\caption{Average energy consumption}
				\end{subfigure}				
				\caption{Vary the UAV's return speed.} 
				\label{fig:vary_speed}
			\end{figure*}	
%			To evaluate the convergence and optimal policy obtained by our proposed algorithms, we use parameters provided in Table.~\ref{table:simulation_parameters}. 				
			In Fig.~\ref{fig:convergene_policy} (a), we compare the convergence of proposed algorithms in terms of average rewards.		
				At the beginning of the learning processes, the average rewards of proposed approaches are close to each other, approximately $0.35$.
				However, only after $4,000$ iterations, D3QL-TL's average reward is nearly $170\%$ greater than those of other approaches.
				Then, D3QL-TL almost converges to the optimal policy after $7.5$$\times$$10^4$ iterations, and its average reward becomes stable at around $0.78$, which is more than $190\%$ greater than those of other learning algorithms.
				Interestingly, D3QL and D3QL-NoRA converge to policies that achieve similar average rewards.
				This suggests that D3QL is unable to take advantage of the battery replacement action.
				In other words, D3QL cannot effectively handle this complicated decision-making situation.			
				This result implies the outperformance of our proposed algorithm, i.e., D3QL-TL, compared with other methods when addressing the extremely complex problem as the one considered in this paper.	
			
			Next, we show the policy obtained by D3QL-TL after $1.5$$\times$$10^5$ iterations in Fig.~\ref{fig:convergene_policy} (b).
				In particular, a point indicates the energy level of the UAV at the beginning of a time slot. 
				The slope of a straight line between two points indicates the selected action, e.g., the steeper this line is, the higher speed is selected. 
				Generally, the lowest speed is selected in a zone that has a high probability of successfully collecting a packet. 
				In contrast, the highest speed is selected in a zone that has a low probability of receiving a packet, as shown in Fig.~\ref{fig:convergene_policy} (b).
				Given $\mathbf{p} = [0.1,0.25,0.6,0.15]$, as in Table.~\ref{table:simulation_parameters}, the D3QL-TL selects the lowest speed in zone $3$, and the highest speed in zone $1,2,$ and $4$.
				More interestingly, the UAV experiences the middle and high speeds in zone $1$. 
				In particular, it travels at the highest speed until its energy decreases to $55$ energy units at time slot $88$, then the middle speed is selected.
				When its energy drops to $40$ energy units at time slot $93$, equivalent to $13.3\%$ energy level, the UAV takes the battery replacement action. 
				Note that Fig.~\ref{fig:convergene_policy} (b) also reveals information of the location where the UAV should take the battery replacement action by energy replenishment time, i.e., $t_{e}$.
				Particularly, given $t_{e}=12$ time slots, the location that the UAV decides to return only $1m$ away from the station.
				This result demonstrates the impacts of location and energy level on the optimal policy of the UAV.
		
		\subsubsection{Performance Evaluation}
		In this section, we perform simulations to evaluate our proposed algorithms in terms of average reward, throughput, and system energy consumption.			
			The parameters are set to be the same as those in~\ref{subsubsec:convergence}.
			The policies of proposed learning algorithms, including Q-learning, D3QL, and D3QL-TL, are obtained after $1.5$$\times$$10^5$ iterations.
							
			In Fig.~\ref{fig:vary_battery_replacement}, we vary the battery replacement time, i.e., $t_{b}$. 				
				Clearly, the average reward and throughput of all policies decrease as the battery replacement time increases from $5$ to $50$ time slots.
				This is stemmed from the fact that given a fixed duration, the less time the UAV needs to replace the battery, the more time it can spend collecting data.
				As a result, the data collection efficiency of the system reduces.
				It can be observed that D3QL-TL can significantly outperform other approaches in terms of average reward and throughput, while it still obtains a reasonable energy consumption per time slot.
				In particular, the average reward and throughput achieved by D3QL-TL are up to $200\%$ and $185\%$ greater than those of the second-best policy, i.e., D3QL, respectively.
				%from 199\%  to 169 \% %from 185\% to 161\%				
							
				\begin{figure*}
					\centering
					\begin{subfigure}[b]{0.3\textwidth}
						\centering
						\includegraphics[scale=0.2]{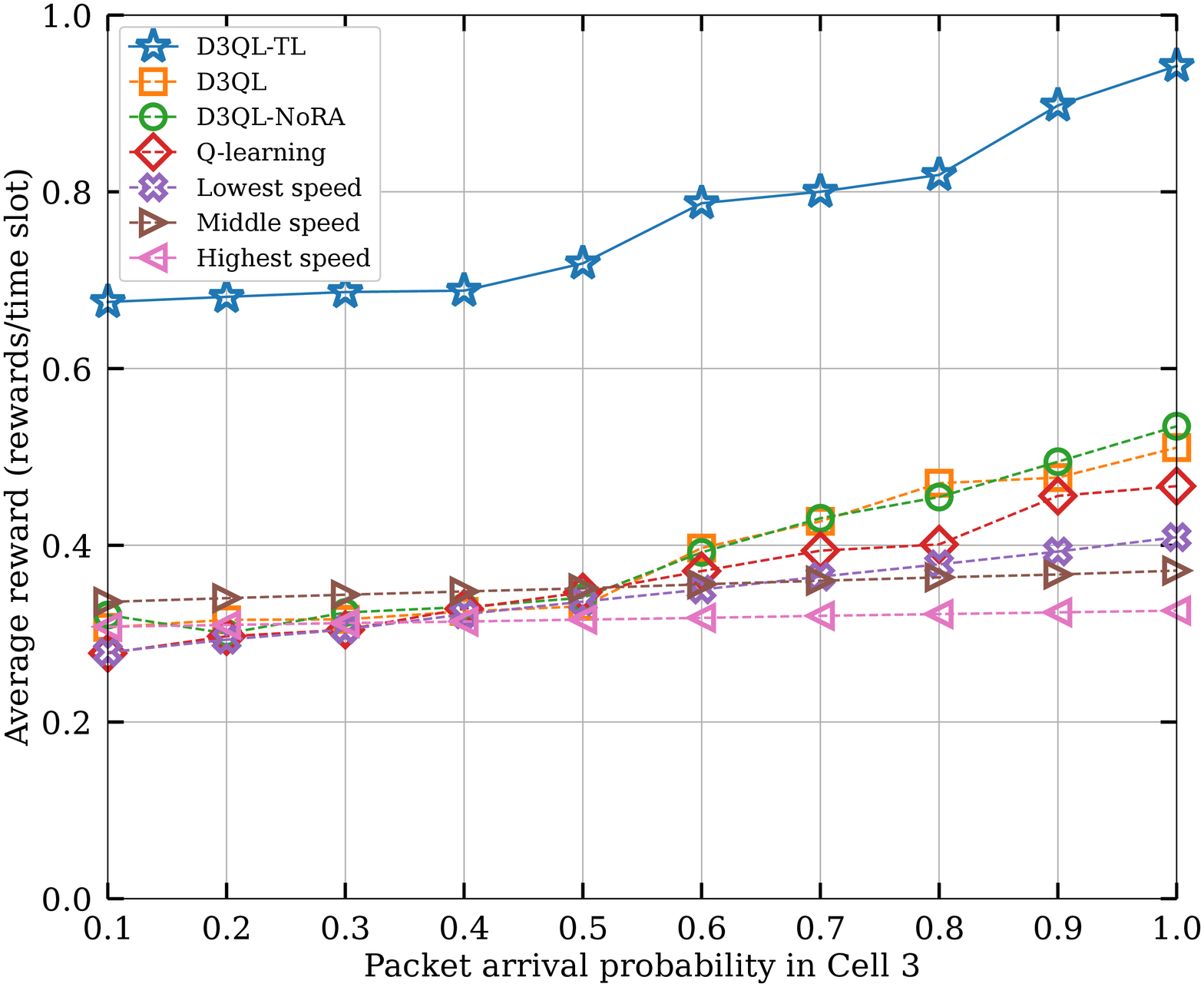}
						\caption{Average reward}
					\end{subfigure}%
					~ 
					\begin{subfigure}[b]{0.3\textwidth}
						\centering
						\includegraphics[scale=0.2]{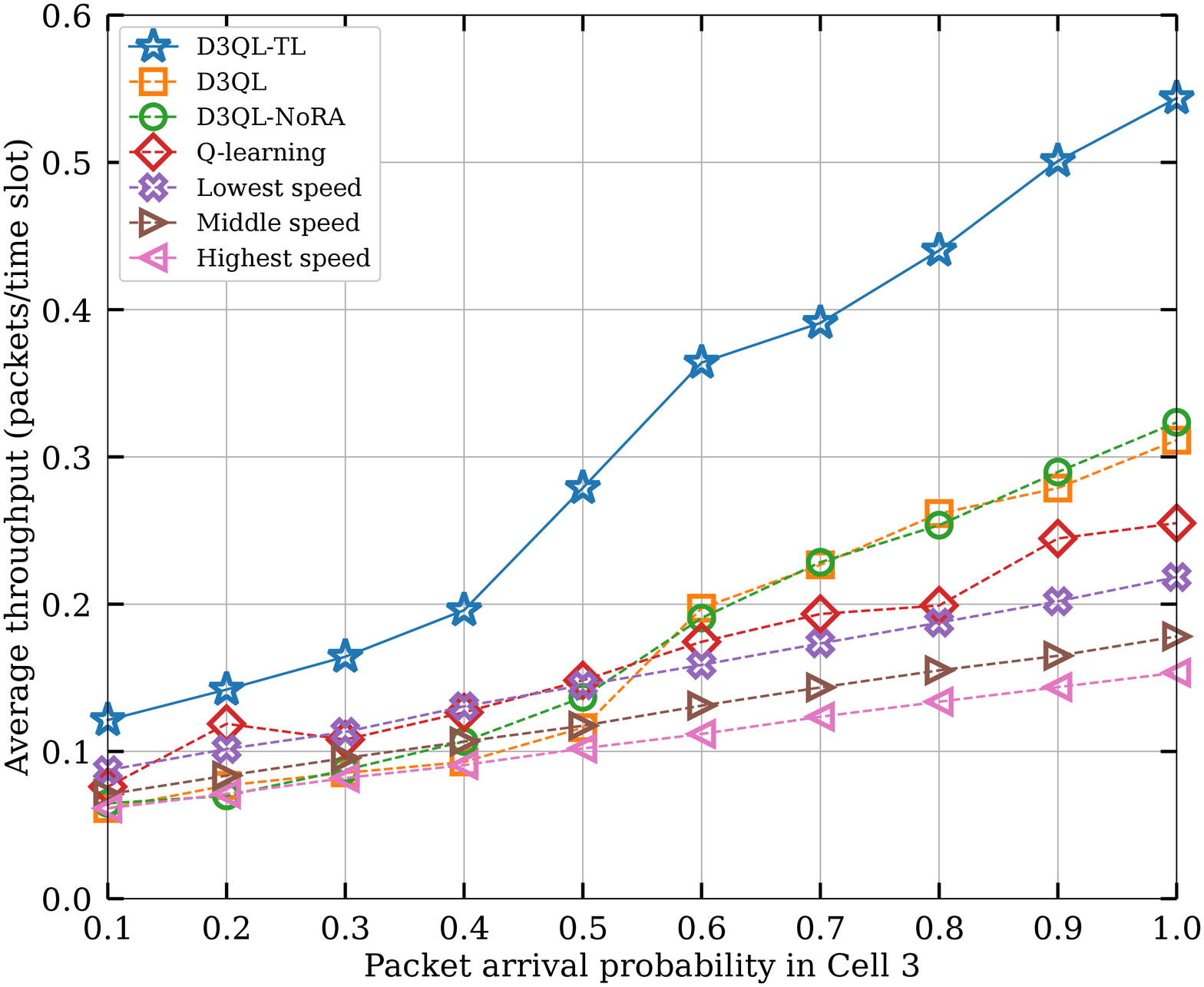}
						\caption{Average throughput}
					\end{subfigure}
					~
					\begin{subfigure}[b]{0.3\textwidth}
						\centering
						\includegraphics[scale=0.2]{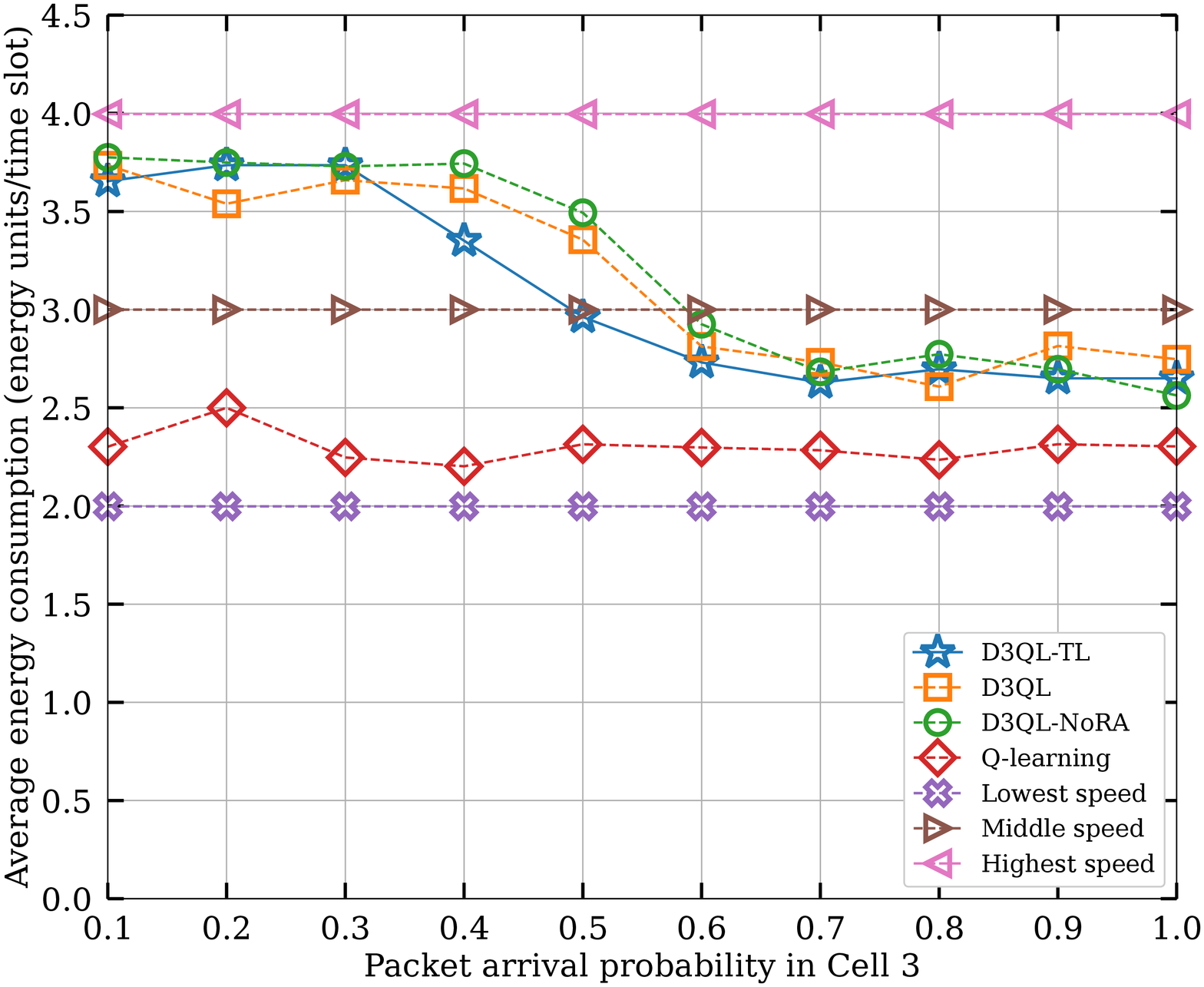}
						\caption{Average energy consumption}
					\end{subfigure}
					
					\caption{Vary the packet arrival probability of zone $3$.} 
					\label{fig:vary_prob}
				\end{figure*}
			
				\begin{figure*}
					\centering
					\begin{subfigure}[b]{0.3\textwidth}
						\centering
						\includegraphics[scale=0.2]{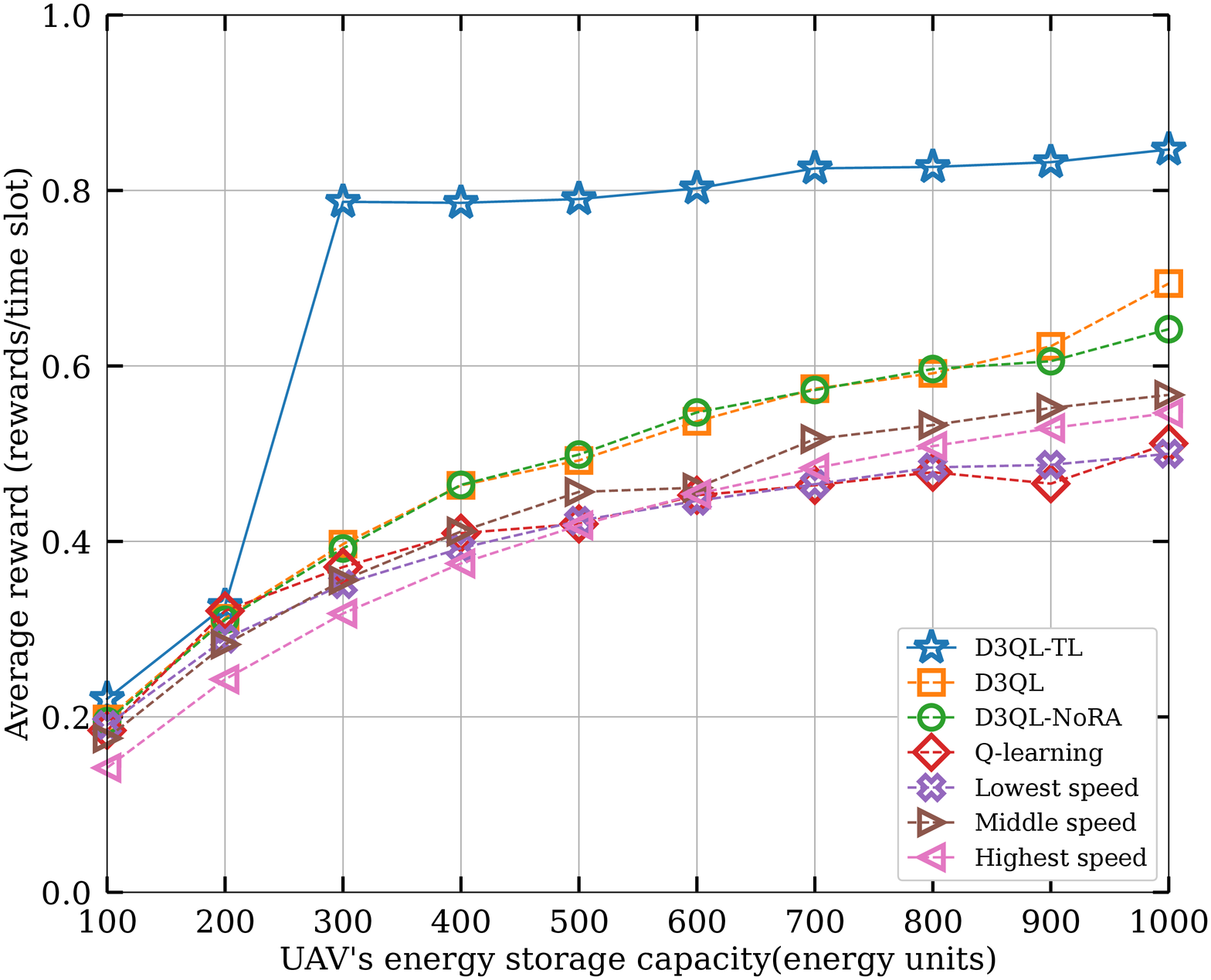}
						\caption{Average reward}
					\end{subfigure}%
					~ 
					\begin{subfigure}[b]{0.3\textwidth}
						\centering
						\includegraphics[scale=0.2]{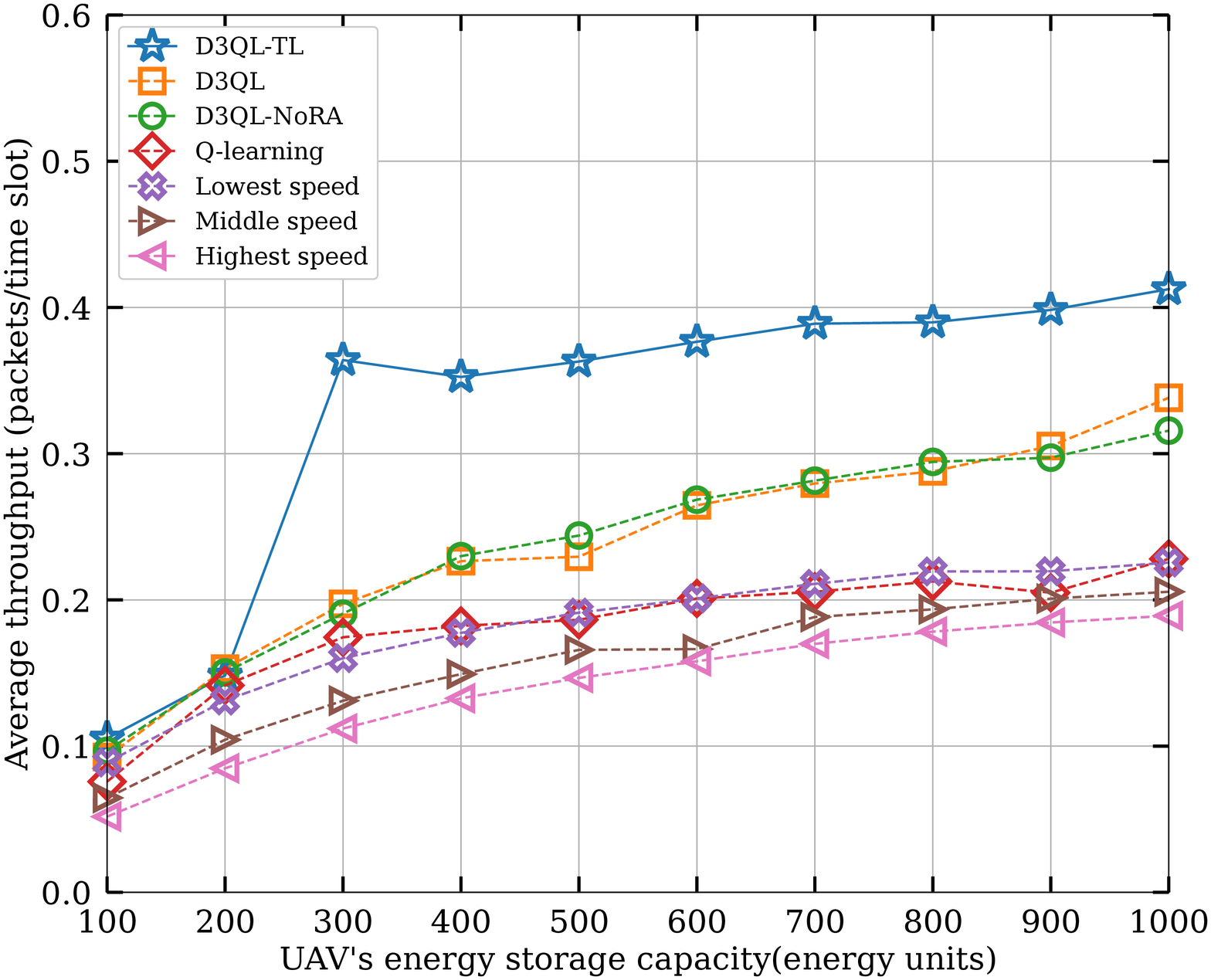}
						\caption{Average throughput}
					\end{subfigure}
					~
					\begin{subfigure}[b]{0.3\textwidth}
						\centering
						\includegraphics[scale=0.2]{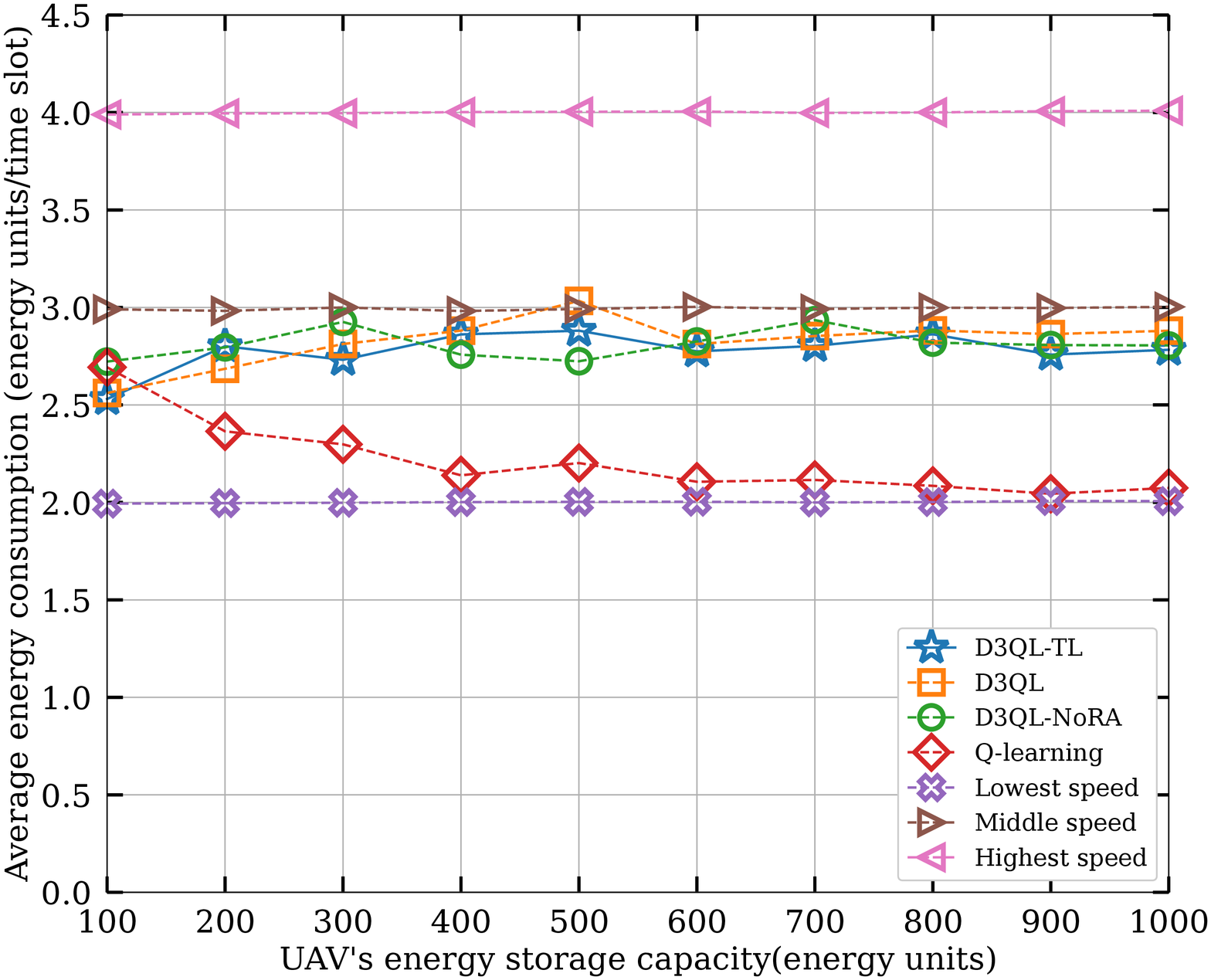}
						\caption{Average energy consumption}
					\end{subfigure}
					
					\caption{Vary the UAV's energy capacity.} 
					\label{fig:vary_energy}
				\end{figure*}
				
				\begin{figure}
					\centering
					$\begin{array}{c}
						\includegraphics[width=0.7\linewidth]{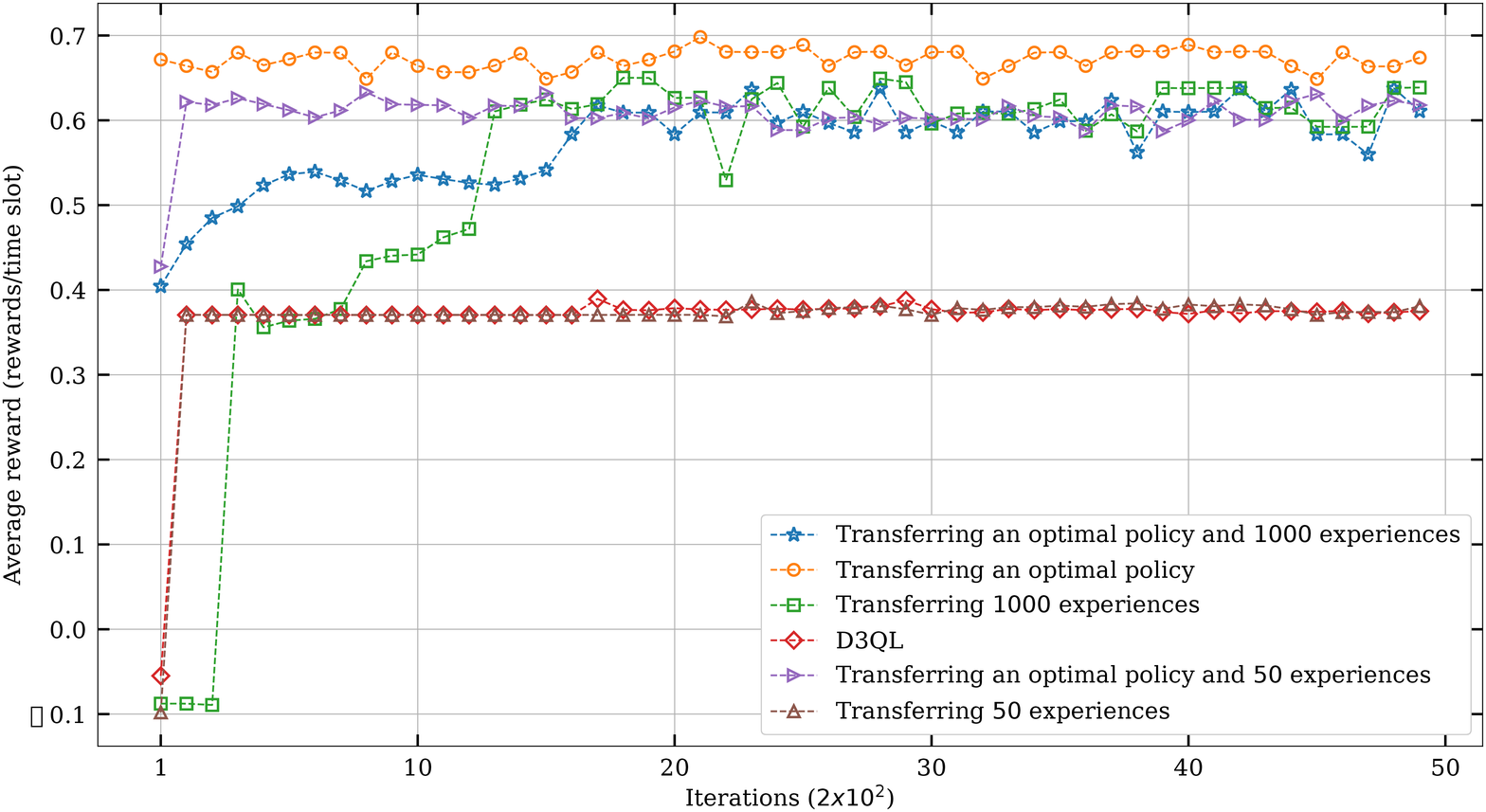}	\\						\text{(a) The first scenario} \\
						\includegraphics[width=0.7\linewidth]{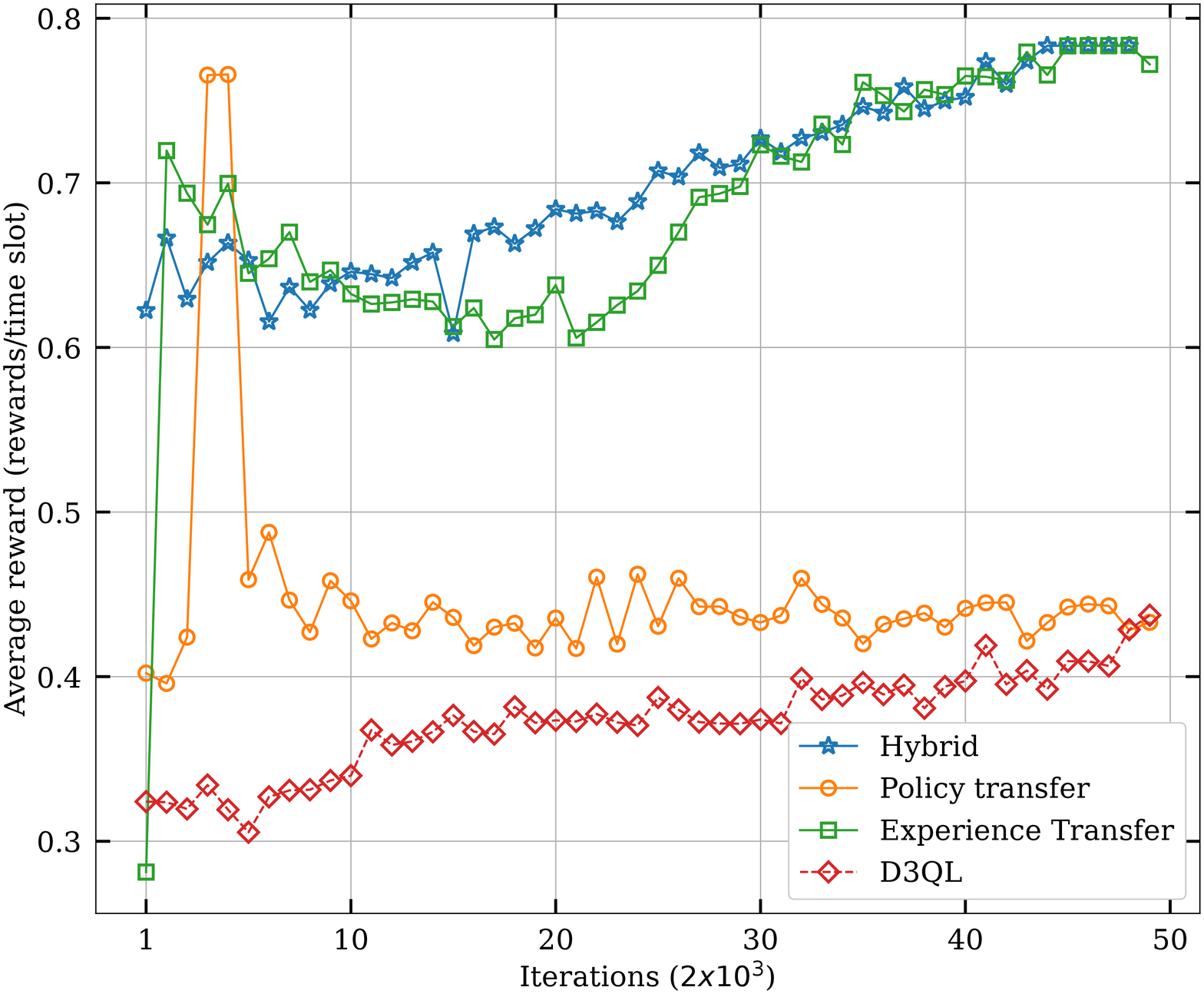} \\
						\text{(b) The second scenario} 
					\end{array}$
					\caption{Convergence of the proposed TL schemes.}
					\label{fig:tl_convergence}
				\end{figure}	
			Next, we vary the return speed $v_r$ of the UAV to observe the system's performance in terms of average reward, throughput, and energy consumption.
				Figs.~\ref{fig:vary_speed} (a) and (b) show that as the return speed $v_r$ increases from $1(m/s)$ to $10(m/s)$, all policies have upward trends in terms of the average reward and throughput, except that of the D3QL-TL.
				The reason is that the increase of $v_r$ leads to the decrease of time for returning to the station, i.e., $t_f$, and thus the UAV has more time to collect data in a fixed duration.  
				More interestingly, when the return speed is low (e.g., lower than $3(m/s)$), the lowest speed policy obtains a higher average reward than that of the highest speed policy, as observed in Fig.~\ref{fig:vary_speed} (a). 
				Nevertheless, the lowest speed policy obtains the lowest performance (i.e., the average reward is lowest)  when the return speed is large.
				This emanates from the fact that given a fixed serving time, the energy consumption of the highest speed is higher than that of the lowest speed.
				Therefore, the UAV has to replace its battery more frequently if it flies at the highest speed rather than if it flies at the lowest speed. 
				Consequently, as the return speed increases, the highest speed policy eventually performs better than that of the lowest speed policy.
				Unlike other policies, D3QL-TL achieves a stable average reward, approximately $0.8$, that is always much higher than those of other policies, as shown in Fig.~\ref{fig:vary_speed} (a).
				This is because the UAV equipped with D3QL-TL can learn an excellent policy, e.g., taking battery replacement action when the UAV is close to the station, making it more adaptable to the changes of the return speed.
				
			We then vary the packet arrival probability $p_3$ of zone $3$ while those of other zones are unchanged, as provided in Table~\ref{table:simulation_parameters}, and observe the performance of our proposed approaches.
				Figs.~\ref{fig:vary_prob} (a) and (b) clearly show the increase of average rewards and throughputs for all policies when $p_3$ increases from $0.1$ to $1.0$.
				Interestingly, as shown in Fig.~\ref{fig:vary_prob} (a), when $p_3$ is small, e.g., lower than $0.4$, the lowest speed policy obtains lower rewards than those of the highest speed.
				However, when $p_3$ becomes larger, the lowest speed achieves a higher average reward than that of the highest speed.
				This implies that the UAV should fly at the lowest speed if the packet arrival probability is high, and vice versa.				
				Fig.~\ref{fig:vary_prob} (c) demonstrates that our proposed algorithm, i.e., D3QL-TL, can learn the environment's dynamic, e.g., the packet arrival probability.
				In particular, when the probability of receiving a packet is low, e.g., less than $0.4$, the UAV's average energy consumption is high, approximately $3.65$ energy units/time slot, indicating that the highest speed is selected more frequently than the lowest speed. 
				In contrast, when this probability becomes higher, e.g., larger than $0.5$, the UAV's energy consumption decreases to around $2.6$, implying that the lowest speed is the most frequently selected speed.
				To that end, D3QL-TL can leverage this knowledge to consistently obtain the best performance compared with other policies.
				 			
			Finally, in Fig.~\ref{fig:vary_energy}, we vary the UAV's energy storage capacity $E$ to study its impact on the system performance. 
				In particular, when $E$ is varied from $100$ to $1000$ energy units, the average rewards and throughputs of all policies increase, as shown in Figs.~\ref{fig:vary_energy} (a) and (b), respectively.
				It is worth highlighting that if $E$ is small, e.g., less than $500$, the performance of the lowest speed is better than that of the highest speed, as illustrated in Fig.~\ref{fig:vary_energy}~(a). 
				However, the highest speed outperforms the lowest speed when $E$ is larger than $500$. 
				This due to the fact that when $E$ is small, the UAV's battery has to be replaced more frequently, leading to a downgrade of the UAV's data collection efficiency.
				Thus, the UAV must conserve more energy by flying at the lowest speed.
				By balancing between energy usage efficiency and data collection efficiency, our proposed D3QL-TL approach can always achieve the highest performance compared to other policies.	
		
		\subsubsection{Transfer Learning Strategies}
		
		In this section, we evaluate and compare the effectiveness of three TL types in D3QL-TL (i.e., experience transfer (ET), policy transfer (PT), and hybrid transfer) in different scenarios, as shown in Figs.~\ref{fig.simulation_MDPs} (b) and (c).
			In particular, the source MDP, i.e., $\mathcal{M}_\mathrm{S}$ is defined as the MDP described in Section~\ref{subsec:PS}, and the simulation parameters are also provided in Table.~\ref{table:simulation_parameters}. 
			Then, the optimal policy obtained by D3QL-TL after $1.5$$\times$$10^5$ iterations and the UAV's experiences gathered in the source MDP are leveraged to reduce the learning time and learning quality.
			To gain an insight of when and how much these transfer learning techniques can improve the learning process of D3QL, we define two scenarios as follows:
			\begin{itemize}
				\item In the first scenario (illustrated in Fig.~\ref{fig.simulation_MDPs} (a)), the target MDP (i.e., $\mathcal{M}_\mathrm{T}^1$) is the same as $\mathcal{M}_\mathrm{S}$ except the trajectory. The UAV only flies over two zones, i.e., zone $1$ and $2$. In each zone, it travels $80 m$.
				\item In the second scenario (illustrated in Fig.~\ref{fig.simulation_MDPs} (b)), the difference between target MDP (i.e., $\mathcal{M}_\mathrm{T}^2$) and $\mathcal{M}_\mathrm{S}$ is the probabilities of receiving a packet, i.e., $\mathbf{p}=[0.6, 0.15, 0.1, 0.25]$.
			\end{itemize}
		We first compare the convergence rate of several transfer learning schemes of D3QL-TL and D3QL in the first scenario in Fig.~\ref{fig:tl_convergence} (a). 
			To investigate how experiences impact on the learning process of D3QL-TL, we select two sizes of experience sets, which are $50$ and $100$.
			As shown in Fig.~\ref{fig:tl_convergence} (a), after $5$$\times$$10^4$ learning iterations, the average rewards obtained by all types of D3QL-TL can achieve up to $179\%$ greater than that of D3QL, except the ET with the size of $50$.  
			For the ET group, when the experience size is small, e.g., $50$, transfer learning does not improve the learning process of D3QL since the average reward of ET is similar to that of D3QL.
			However, when this size is large enough, e.g., $10^3$, the asymptotic performance of D3QL-TL is $171\%$ greater than that of D3QL in terms of average reward.
			Interestingly, for the hybrid approach, when the experience size decreases, the D3QL-TL's performance increases.
			Especially, the PT, which is equivalent to the hybrid with zero experience size, consistently outperforms other transfer approaches in all the metrics, i.e., jump-start, time-to-threshold, and asymptotic performance. 
			Specifically, only after $10^3$ iterations, the PT obtains the optimal policy, and its average reward is stable at around $0.68$.
			These results demonstrate that if the environment's dynamics in the target MDP, e.g., probabilities of receiving a packet, are similar to those in the source MDP, PT is the best choice.
			The reason is that the change of trajectory makes source experiences less efficient than that of the source policy. 
			It is worth noting that only schemes with PT can improve the system performance at the beginning of the learning process because this policy can help the UAV choose valuable actions in this period, e.g., selecting battery replacement action when it is near the station and its energy level is low.
						
		 In Fig.~\ref{fig:tl_convergence} (b), we show the results of the second scenario where the packet arrival probabilities are different from that of the source MDP.
			 In this scenario, we set the experience size to $10^3$ for the ET and hybrid schemes.
			 Again, it can be observed that only schemes with PT can improve the initial performance, e.g., during the first $2$$\times$$10^3$ iterations.
			 Unlike the first scenario, the PT yields the worst performance among the transfer learning schemes, and its asymptotic performance is almost zero, meaning that there is no improvement in terms of average reward at the end of the learning process.
			 In contrast, the ET and hybrid schemes achieve similar asymptotic performance, approximately $172\%$.
			 However, hybrid's jump-start metric, i.e., $192\%$, is significantly higher than that of ET, i.e., $86\%$.
			 These results suggest that when the environment dynamics change, the hybrid scheme should be selected.

\section{Conclusions}
\label{sec:conclusions}
	In this paper, we develop a novel Deep Dueling Double Q-learning with Transfer Learning algorithm (D3QL-TL) that jointly optimizes the flying speed and energy replenishment activities for the UAV to maximize the data collection performance of a UAV-assisted IoT system.
	The proposed algorithm effectively addresses not only the dynamic and uncertainty of the system but also the high dimensional state and action spaces of the underlying MDP problem with hundreds of thousands of states.
	In addition, the proposed TL techniques (i.e., experience transfer, policy transfer, and hybrid transfer) allow UAVs to ``share'' and ``transfer'' their learned knowledge, resulting in a decrease of learning time and an improvement of learning quality.
	The simulation results show that our proposed solution can significantly improve the system performance (i.e., data collection and energy usage efficiency) compared with other conventional approaches.

%	In this paper, we have introduced a novel framework that jointly optimizes the flying speed and energy replenishment activities for the UAV to maximize the data collection performance of a UAV-assisted IoT system. 
%	In particular, to capture the dynamics and uncertainty of the system, a Markov decision process has been developed, taking into account the UAV's location and energy level. To address the curse-of-dimensionality problem of conventional Q-learning-based algorithms, we have proposed a highly effective algorithm, namely D3QL, by utilizing the recent advances in deep reinforcement learning and deep dueling network architecture. After that, we have introduced three transfer techniques, including experience transfer, policy transfer, and hybrid transfer, to further improve and stable the learning process of the D3QL algorithm. The simulation results showed that our proposed solution can significantly improve the system performance compared with other conventional approaches.
%{0.7cm}
%--------------------------------------------------------------------------------------------------------------------------
%--------------------------------------------------------------------------------------------------------------------------	
\section*{Acknowledgment}
The first author is supported by the Vingroup Science and Technology Scholarship Program for Overseas Study for Master's and Doctoral Degrees.

\renewcommand{\baselinestretch}{1}


\begin{thebibliography}{1}

\bibitem{chu_fast_2021}
N.~H.~Chu, D.~T.~Hoang, D.~N.~Nguyen, N.~V.~Huynh and E.~Dutkiewicz, ``Fast or slow: An autonomous speed control approach for UAV-assisted IoT data collection networks,'' in \emph{Proceedings of 2021 IEEE Wireless Communications and Networking Conference (WCNC),} 2021, pp. 1-6.

\bibitem{Cisco}
Cisco Annual Internet Report (2018-2023) White Paper, [Online]. Available: https://www.cisco.com/c/en/us/solutions/collateral/executive-perspectives/annual-internet-report/white-paper-c11-741490.html

\bibitem{Wang2019Joint}
J.~Wang, C.~Jiang, Z.~Wei, C.~Pan, H.~Zhang, and Y.~Ren, ``Joint UAV hovering altitude and power control for space-air-ground IoT networks,'' \emph{IEEE Internet of Things Journal}, vol. 6, no. 2, pp. 1741-1753, Apr. 2019.

\bibitem{Liu2020Distributed}
C.~H.~Liu, X.~Ma, X.~Gao, and J.~Tang, ``Distributed energy-efficient multi-UAV navigation for long-term communication coverage by deep reinforcement learning,'' \emph{IEEE Transactions on Mobile Computing}, vol. 19, no. 6, pp.~1274-1285, Jun.~2020.

\bibitem{facebook_connecting_2014}
Facebook,  ``Connecting  the  world  from  the  sky,''  \emph{Facebook  Technical Report,} 2014.  

\bibitem{google_loon}
https://loon.com/

\bibitem{shan_looking_2020}
F.~Shan, J.~Luo, R.~Xiong, W.~Wu, and J.~Li, ``Looking before crossing: An optimal algorithm to minimize UAV energy by speed scheduling with a practical flight energy model,'' in \emph{Proceedings of the 2020 IEEE Conference on Computer Communications (INFOCOM),} IEEE, 2020, pp.~1758-1767.

\bibitem{zeng_age_2020}
X. Zeng, F. Ma, T. Chen, X. Chen and X. Wang, ``Age-optimal UAV trajectory planning for information gathering with energy constraints,'' in \emph{Proceedings of 2020 IEEE/CIC International Conference on Communications in China (ICCC),} 2020, pp.~881-886.


\bibitem{bouhamed_uav_2020}
O. Bouhamed, H. Ghazzai, H. Besbes and Y. Massoud, ``A UAV-assisted data collection for wireless sensor networks: Autonomous navigation and scheduling,'' \emph{IEEE Access,} vol. 8, pp.~110446-110460, Jun.~2020.


\bibitem{fu_energy_2021}
S. Fu et al., ``Energy-efficient UAV-enabled data collection via wireless charging: A reinforcement learning approach,'' \emph{IEEE Internet of Things Journal,} vol. 8, no. 12, pp. 10209-10219, Jun.~2021.

\bibitem{zhang_hierarchical_2021}
Y. Zhang, Z. Mou, F. Gao, L. Xing, J. Jiang and Z. Han, ``Hierarchical deep reinforcement learning for backscattering data collection with multiple UAVs,'' \emph{IEEE Internet of Things Journal,} vol. 8, no. 5, pp. 3786-3800, Mar.~2021.

\bibitem{xu_blockchain_2021}
X. Xu, H. Zhao, H. Yao and S. Wang, ``A blockchain-enabled energy-efficient data collection system for UAV-assisted IoT,'' \emph{IEEE Internet of Things Journal,} vol. 8, no. 4, pp. 2431-2443, Feb.~2021.
%%%%%%%%%%%%%%%%%%%

%%% speed control
\bibitem{Gong2018JSAC}
J. Gong, T. Chang, C. Shen and X. Chen, ``Flight time minimization of UAV for data collection over wireless sensor networks," \emph{IEEE Journal on Selected Areas in Communications}, vol. 36, no. 9, pp.~1942-1954, Sep.~2018.

\bibitem{Pan2018Sensor}
Q. Pan, X. Wen, Z. Lu, L. Li, and W. Jing, ``Dynamic speed control of unmanned aerial vehicles for data collection under internet of things,'' \emph{Sensors}, vol. 18, no. 11, Nov.~2018.	

\bibitem{Lin2019JIoT}
X. Lin, G. Su, B. Chen, H. Wang and M. Dai, ``Striking a balance between system throughput and energy efficiency for UAV-IoT systems,'' \emph{IEEE Internet of Things Journal}, vol. 6, no. 6, pp.~10519-10533, Dec.~2019.	

\bibitem{li_onboard_2019}
K. Li, W. Ni, E. Tovar and A. Jamalipour, ``On-board deep Q-network for UAV-assisted online power transfer and data collection,'' \emph{IEEE Transactions on Vehicular Technology,} vol. 68, no. 12, pp. 12215-12226, Dec. 2019.

\bibitem{Wu2018OFDMA}
Q. Wu and R. Zhang, ``Common throughput maximization in UAV-enabled OFDMA systems with delay consideration,'' \emph{IEEE Transactions on Communications}, vol. 66, no. 12, pp.~6614-6627, Dec.~2018.

\bibitem{elmagid_deep_2019}
M.~A.~Abd-Elmagid, A.~Ferdowsi, H.~S.~Dhillon, and W.~Saad, ``Deep reinforcement learning for minimizing age-of-information in UAV-assisted networks.'' in \emph{Proceedings of the 2019 IEEE Global Communications Conference (GLOBECOM),} IEEE, 2019, pp.~1-6.

%%%%%%%%%%%%%%
\bibitem{Hasselt2016}
H.~Hasselt, A.~Guez, and D.~Silver, ``Deep reinforcement learning with double Q-learning." in \emph{Proceedings of the Thirtieth AAAI Conference on Artificial Intelligence (AAAI'16)}, AAAI Press, 2016, pp.~2094-2100.

\bibitem{Wang2016Dueling}
Z.~Wang et al., ``Dueling network architectures for deep reinforcement learning,'' in \emph{Proceedings of The 33rd International Conference on Machine Learning}, 2016, pp. 1995-2003. 

\bibitem{khodr_energy_2017}
H. Khodr, N. Kouzayha, M. Abdallah, J. Costantine and Z. Dawy, ``Energy efficient IoT sensor with RF wake-up and addressing capability,'' \emph{IEEE Sensors Letters,} vol. 1, no. 6, pp. 1-4, Dec. 2017.

\bibitem{du_gradient_2019}
S.~Du, J.~Lee, H.~Li, L.~Wang, and X.~Zhai, ``Gradient descent finds global minima of deep neural networks,'' in \emph{Proceedings of the 36th International Conference on Machine Learning, in Proceedings of Machine Learning Research,} 2019, pp. 1675-1685.

\bibitem{Robbins1951SGD}
H.~Robbins and S.~Monro, ``A stochastic approximation method,'' \emph{Ann. Math. Stat.}, vol. 22, no. 3, pp. 400-407, 1951.

\bibitem{Watkins1992QLearning}
C.~J.~C.~H.~Watkins and P.~Dayan, ``Q-learning,'' \emph{Mach. Learn.}, vol. 8, no. 3-4, pp. 279-292, 1992.

\bibitem{Sutton1998Reinforcement}
R.~S.~Sutton and A.~G.~Barto, \emph{Reinforcement Learning: An Introduction}. Cambridge, MA, USA: MIT Press, 1998.

\bibitem{Mnih2015Human}
V.~Mnih et al., ``Human-level control through deep reinforcement learning,'' \emph{Nature}, vol. 518, no. 7540, pp. 529-533, Feb. 2015.

\bibitem{Thrun1993}
S.~Thrun and A.~Schwartz, ``Issues in using function approximation for reinforcement learning,'' in \emph{Proceedings of the 1993 Connectionist Models Summer School}, Hillsdale, NJ, 1993.

\bibitem{Goodfellow2016Deep}
I.~Goodfellow, Y.~Bengio, and A.~Courville, \emph{Deep learning}. MIT press, 2016.

\bibitem{halkjaer_the_1996}
S.~Halkjear and O.~Winther, ``The effect of correlated input data on the dynamics of learning.'' in \emph{Proceedings of the 9th International Conference on Neural Information Processing Systems}, 1996, pp. 169-175.

\bibitem{taylor_transfer_2009}
M.~E.~Taylor and P.~Stone. ``Transfer learning for reinforcement learning domains: A survey.'' \emph{Journal of Machine Learning Research}, no. 7, pp. 1633-1685, 2009.

\bibitem{pan_survey_2009}
S. J. Pan and Q. Yang, ``A Survey on Transfer Learning,'' \emph{IEEE Transactions on Knowledge and Data Engineering,} vol. 22, no. 10, pp.~1345-1359, Oct.~2009.


\bibitem{zhu_trasfer_2020}
Z.~Zhu, K.~Lin, and J.~Zhou, ``Transfer learning in deep reinforcement learning: A survey.'' arXiv:2009.07888 [cs.LG], Sep.~2020.

\bibitem{nguyen_transfer_2021}
C.~T.~Nguyen, N.~V.~Huynh, N.~H.~Chu, Y.~M.~Saputra, D.~T.~Hoang, D.~N.~Nguyen, Q.~V.~Pham, D.~Niyato, E.~Dutkiewicz, and W.~.J~Hwang, ``Transfer learning for future wireless networks: A comprehensive survey.'' arXiv:2102.07572v1 [cs.LG], Feb.~2021. 

\end{thebibliography}
\end{document}